\input jnl.tex
\input reforder.tex
\input epsf.tex

\def\a{{\alpha}}

\def\d{{\partial}}
\def\e{{\epsilon}}
\def\D{{\cal D}}

\def\half{{1 \over 2}}
\def\ra{{\rangle}}
\def\la{{\langle}}

\def\ih{{i \over \hbar}}

\def\E{{\cal E}}
\def\au{\underline \alpha}
\def\p{{\bf p}}
\def\q{{\bf q}}
\def\r{{\bf r}}

\def\v{{\bf v}}
\def\x{{\bf x}}
\def\X{{\bf X}}
\def\y{{\bf y}}

\def\ria{{\rightarrow}}

\centerline{\bf Life in an Energy Eigenstate:}
\centerline{\bf Decoherent Histories Analysis of
a Model Timeless Universe.}
\vskip 0.3in
\author J.J.Halliwell % \footnote{$^{\dag}$}{E-mail address: \jjh}
\vskip 0.2in
\centerline{\rm and}
\vskip 0.2in
\author J.Thorwart

\vskip 0.3in
\affil
Theory Group
Blackett Laboratory
Imperial College
London SW7 2BZ
UK
\vskip 0.5in
\centerline {\rm Preprint IC/TP/1-02/13. January, 2002}
\vskip 0.1in
%\vskip 1.0in
%\centerline {\rm  PACS numbers: 04.60.-m, 04.60.Gw, 04.60.Kz, 98.80.Hw}
%\centerline{\bf NOT FOR CIRCULATION. PRELIMINARY DRAFT}

\vskip 0.1 in

\abstract {Inspired by quantum cosmology, in which the wave
function of the universe is annihilated by the total Hamiltonian,
we consider the internal dynamics of a simple particle system in
an energy eigenstate. Such a system does not possess a uniquely
defined time parameter and all physical questions about it must be
posed without reference to time. We consider in particular the
question, what is the probability that the system's trajectory
passes through a set of regions of configuration space without
reference to time? We first consider the classical case, where
the answer has a variety of forms in terms of a phase space
probability distribution function. We then consider the quantum
case, and we analyze this question using the decoherent histories
approach to quantum theory, adapted to questions which do not
involve time. When the histories are decoherent, the probabilities
approximately coincide with the classical case, with the phase
space probability distribution replaced by the Wigner function of
the quantum state. For some initial states, decoherence requires
an environment, and we compute the required influence functional
and examine some of its properties. Special attention is given to
the inner product used in the construction (the induced or Rieffel
inner product), the construction of class operators describing the
histories, and the extent to which reparametrization invariance is
respected. Our results indicate that simple systems without an
explicit time parameter may be quantized using the decoherent
histories approach, and the expected classical limit extracted.
The results support, for simple models, the usual heuristic
proposals for the probability distribution function associated
with a semiclassical wave function satisfying the Wheeler-DeWitt equation. }

\endtopmatter
\endpage

\head{\bf 1. Introduction}

There are a variety of interesting physical situations in which
our classical view of the world inspires us to ask questions that
quantum theory does not easily answer. For example, classical
mechanics concerns simultaneously specified values of coordinates
and momenta while quantum theory has to go through some
contortions to say what this means operationally. Another
important class of problems of this type are those that involve
time in a non-trivial way. The arrival time problem and tunneling
time problem, for example, have been the subject of considerable
recent interest [\cite{Time}].

Perhaps more intriguing than these are questions in quantum
mechanics that do not involve time at all. Consider for example
the following situation. Suppose we have a system of particles in
a state of fixed total energy. It could, for example, be a light
particle orbiting a massive particle. Then classically, we can ask
whether the light particle passes through a certain region of
configuration space at any stage during its orbit. Or we can ask
which of the possible classical orbits the light particle follows.
The important point here is that these sorts of questions do not
involve time explicitly. Much experimental and observation data is
in fact of this type. For example, astronomical observations yield
planetary orbits, and particle physics experiments often yield a
photograph of a track in a bubble chamber. We would like to ask
the same question in quantum theory: given that the system is in
an energy eigenstate, what is the probability that it is found in
a certain region of configuration space, irrespective of time?

The primary motive for considering this question is
quantum cosmology [\cite{HaHa}]. There, in simple cosmological models,
the wave function of the system obeys the Wheeler-DeWitt
equation,
$$
H \Psi ( \x) = 0
\eqno(1.1)
$$
where $H$ is the total Hamiltonian of the gravitational field plus
all matter sources in the universe. The most significant feature
of this equation is that it contains absolutely no reference to
time whatsoever. It is usually argued that ``time'', or more
precisely, the physical systems that we use to measure time, are
contained already in the gravitational and matter fields
[\cite{Kuc,Ish,BuIs}]. While this is very plausible it leaves us
with the question as to how to extract interesting physical
predictions from this wave function, given the absence of the time
coordinate that plays such a central role in standard quantum
theory. In quantum cosmology, heuristic methods (mainly the ``WKB
interpretation'') have been used to extract useful predictions
[\cite{Hal1}], but what we are concerned with here is how such
heuristic ideas may be incorporated in a properly defined
interpretational framework for the quantum theory of timeless
models. Furthermore, although simple (minisuperspace) models are
unlikely to be reasonable physical approximations to a full
quantum gravity theory, it does seem likely that a quantum theory
of gravity will centre around a timeless equation of the form
(1.1) (the loop variable approach, for example, involves such an
equation [\cite{Ash}]), hence it remains important to understand
the quantization and interpretation of such systems at an
elementary level.

Quantum cosmology, then, is our main motivation for studying the
internal dynamics of a closed system in an energy eigenstate. The
question we shall focus on is this: given an energy eigenstate,
what is the probability that the system enters a region $\Delta $
of configuration space? Or similarly, suppose the coordinates of
the system are $\x = (x_1, x_2, \cdots x_n) $, what is the
probability distribution of, say, $x_2, \cdots x_n$, given the
value of $x_1$? Classically, these questions may be answered
reasonably easily. We look for all the phase space initial data
points whose classical trajectories pass through $\Delta$, and
then the desired probability for entering $\Delta$ is
the probability measure on this subset of phase space. In
quantum theory the question is considerably more complicated, and,
like the question of phase space samplings, a variety of different
(generally inequivalent) approaches may be employed.

It is perhaps of interest to spell out in more detail how the
absence of a time parameter affects these considerations. In a
standard non-relativistic particle theory, the variables of
interest are, for example, positions at a fixed moment of time $\x
(t)$. Here, the time $t$ is regarded as an observable physical
parameter. In a theory without time, by contrast, the quantities
of interest are curves in configuration space $ \x (s)$ (or more
generally, in phase space). Here, $s$ is {\it not} a physically
measurable time, but is simply a parameter labeling the points
along the curve, and the curves are parameterized in this way for
mathematical convenience. Furthermore, one of the characteristic
features of genuinely timeless theories is that none of the
components of $\x$ are monotonic in $s$. This means that it is not
possible to use one of the components of $\x$ as a ``time''
parameter, except for local sections of the curve. Despite these
features, such classical theories are well-defined and predictive.
Indeed, many classical cosmological models are of this type. For
example, in the massive scalar field cosmological model (for a
positive curvature FRW metric) [\cite{Haw}], the classical
solutions go backwards and forwards in both the scale factor $a$
and the scalar field $\phi$. Therefore, the most general such
classical theory is one in which there is a probability
distribution on the set of classical trajectories, and this is
also well-defined (subject to careful normalization) as we shall
see. Such probability distributions are of particular interest for
predicting, for example, the likelihood of the initial conditions
for inflation [\cite{HaPa}].

A closely related issue is the fact that a vanishing Hamiltonian
(or in the quantum theory, Eq.(1.1)) is associated with the
symmetry of reparametrization invariance, which is essentially the
freedom to redefine the parameter $s$ labeling points along a
trajectory. Individual phase space points are not
reparametrization-invariant, since they are moved along the
classical trajectories by a reparametrization. But a useful
invariant quantity is the entire classical trajectory, as we shall
see later in more detail. This simple observation turns out to be
a useful focal point for what we do in this paper.

Turning now to the quantum theory, there are, as stated, many
possible approaches to the these systems, all of which are
complicated by the absence of a time parameter. However, the
aspects of the classical theory outlined above suggest that a
particularly useful approach to quantization is the decoherent
histories approach [\cite{GeHa,Gri,Omn}]. This is because it deals
directly with entire trajectories and does not obviously require a
time coordinate. The aim of this paper is to show that the
decoherent histories approach can be used to calculate
probabilities for histories in configuration space in simple
timeless models. In particular, we shall show that in the
classical limit, the decoherent histories approach produces a set
of classical trajectories with a probability measure on that set.

The other significant approach to this problem that is
currently being pursued involves constructing
operators corresponding to the questions posed above,
and which commute with the Hamiltonian, so are
``observables'' [\cite{DeW,Rov1,Rov2,Rov3,Mar1,Mar2}].
For a free particle in two dimensions, for example, the
classical trajectories have the form,
$$
x_1 (t)  = x_1 + { p_1 t \over m}, \quad
x_2 (t)  = x_2 + { p_2 t \over m}
\eqno(1.2)
$$
and we may eliminate $t$ between them to write,
$$
x_1(t) = x_1 + { p_1 \over p_2} ( x_2 (t) - x_2 )
\eqno(1.3)
$$
This is the classical answer to the question, what
is the value of $ x_1$ at a given value of $x_2$?
It commutes with the free particle Hamiltonian,
$$
H = \half (p_1^2 + p_2^2 )
\eqno(1.4)
$$
so is an observable and we may find states that are eigenstates of
both (1.3) and (1.4), from which one may begin to address the
questions set out above. We will not say much about this approach
here, but it is important to mention because any other approach
ought to make some kind of contact with it at some stage, and in
particular, the decoherent histories approach must contain some
notion corresponding to the notion of an observable in the
operator approach. (See also Refs.[\cite{Har1,Har2}] for some much
earlier approaches to these issues).

We begin Section 2 by describing the classical result. We
introduce a classical phase space distribution function $w(\p,
\x)$ and compute the probability that a trajectory in
configuration space passes through a region $\Delta$. Most of the
key ideas for this paper are in fact contained in this classical
result. In particular, we discuss the reparametrization invariance
of the system, and introduce observables corresponding to entire
classical trajectories. We write the classical result in a number
of different forms, including a form in terms of a flux across a
hypersurface, closely related to the heuristic WKB interpretation
of quantum cosmology.

We begin the quantum case in Section 3, with the construction of
the decoherence functional for timeless models. This has two
important aspects. The first concerns the choice of inner product
in the construction, since solutions to equations of the form
Eq.(1.1) are typically not normalizable in the simple
Schr\"odinger inner product. The appropriate choice is the
induced (or Rieffel) inner product, which we describe. We also
show how this inner product for the quantum case implies
a useful normalization for the classical phase space
distribution function. The second
aspect is the construction of the class operators, which, in this
case, are propagators describing coarse-grained sets of histories
passing through restricted regions of configurations space.

In Section 4, we discuss the semiclassical limit of the
decoherence functional. A key step is the construction of class
operators corresponding to restricted sets of histories entering
the region $\Delta$. The obvious candidates for these class
operators are not in fact compatible with the constraint equation,
and we therefore show how they may be appropriately modified. This
turns out in fact to be the crucial step in the construction of
the decoherence functional. We then show that, for the special
initial states for which the histories are decoherent, the
probabilities for the histories approximately coincide with the
classical case, with the phase space distribution function $w(\p,
\x)$ replaced by the Wigner function $W(\p, \x)$ of the quantum
system.

In Section 5, we consider the special case in which the system is
a collection of harmonic oscillators in a fixed energy eigenstate.
For this system it is possible to introduce a special class of
eigenstates of the Hamiltonian sometimes called ``timeless
coherent states'', which have the property that they are
concentrated about an entire classical phase space trajectory. We
discuss the decoherence and probabilities associated with these
states and obtain the intuitively expected physical results for
the probabilities of entering a region $\Delta$.

Since decoherence is only obtained for special initial states,
we consider, in Section 6, the addition of an environment
to produce decoherence for a wide variety of initial states.
We repeat the calculation of decoherence and probabilities
with intuitively expected results, in agreement with
classical expectations.

The calculations of Sections 4 and 6 used initial states
consisting of single WKB wave packets. In Section 7, we therefore
extend to the case of superpositions of such wave packets. This
turns out in fact to be straightforward, and very similar to
earlier calculations performed with the reduced density matrix. We
easily find that the interference terms between different WKB wave
packets are very small. We summarize and conclude in Section 8.

Finally, we outline the relationship of the present paper to other
works in the field. This paper is part of a programme to apply
the decoherent histories approach to quantum cosmology, or models
of the type used in quantum cosmology, and ultimately to quantum
gravity more generally. The decoherent histories approach
generally and its applications to quantum cosmology have been set
out at length by Hartle in his 1992 Les Houches Lectures
[\cite{Har3}], and more recent relevant aspects of this were
discussed by Hartle and Marolf [\cite{HaMa}]. This formalism have
been applied to particular models in three places, most recently
by the present authors [\cite{HaTh}], who used it to construct in
detail the decoherence functional for the relativistic particle,
and compute probabilities for crossing spacelike surfaces. The
present paper is in some ways an extension of that work. Whelan
[\cite{Whe}] has used the formalism to compute probabilities on
timelike surfaces for the relativistic particle. Also, Craig and
Hartle [\cite{CrHa}] have applied the formalism to a Bianchi IX
quantum cosmological model. The last two papers use the
Klein-Gordon inner product whereas here, we use the
positive-definite induced inner product to construct the
decoherence functional. There is also some connection with the
work on probabilities for non-trivial spacetime coarse grainings
in non-relativistic quantum mechanics [\cite{YaT}].

It is also perhaps worth mentioning that this paper is very much
in the spirit of Ref.[\cite{Hal2}], which attempts to interpret
the Wheeler-DeWitt equation in terms of emergent trajectories, by
introducing model detectors into the Hamiltonian. This was in turn
inspired by Barbour's observations [\cite{Bar1}], on the
similarity between the Wheeler-DeWitt equation and Mott's
calculation showing the emergence of a straight line track from a
spherical wave in alpha decay [\cite{Mot}], together with some of
Barbour's more general observations about the Wheeler-DeWitt
equation and timeless theories [\cite{Bar1,Bar2,Bar3}].

\head{\bf 2. The Classical Case}

We are interested in the question, ``What is the probability
associated with a given region of configuration space when
the system is in an energy eigenstate?". We begin by analyzing
the classical problem.

We will consider a classical system described by a
$2n$-dimensional phase space, with coordinates
and momenta $ (\x, \p ) = (x_k, p_k) $, and Hamiltonian
$$
H = \sum_{k=1}^n \left( {p_k^2 \over 2 M} + V (\x ) \right)
\eqno(2.1)
$$
More generally, we are interested in a system for which the
kinetic part of the Hamiltonian has the form $ g^{kj} (\x) p_k
p_j$, where $g^{kj} (\x)$ is an inverse metric of hyperbolic
signature. Most minisuperspace models in quantum cosmology have a
Hamiltonian of this form. However, the focus of this paper is the
timelessness of the system, and the form of the configuration
space metric turns out to be unimportant. So for simplicity, we
will concentrate on the form Eq.(2.1).

We assume that there is a classical phase space distribution
function $w (\p, \x ) $, which is normalized according to
$$
\int d^n p \ d^n x \ w (\p, \x ) = 1
\eqno(2.2)
$$
and obeys the evolution equation
$$
{ \partial w \over \partial t } = \sum_k
\left( - { p_k \over M} { \partial w \over \partial x_k}
+ { \partial V \over \partial x_k } { \partial w \over \partial p_k
} \right) = \{ H, w \}
\eqno(2.3)
$$
where $\{ \ , \ \}$ denotes the Poisson bracket. The interesting
case is that in which $w$ is the classical analogue of an energy
eigenstate, in which case $ \partial w / \partial t = 0 $, so the
evolution equation is simply
$$
\{ H , w \} = 0
\eqno(2.4)
$$
It follows that
$$
w( \p^{cl} (t), \x^{cl} (t) ) = w ( \p (0), \x (0) )
\eqno(2.5)
$$
where $\p^{cl}(t), \x^{cl}(t)$ are the classical solutions with
initial data $\p(0), \x(0)$, so $w$ is constant along the classical
orbits. (The normalization of $w$ then
becomes an issue if the classical orbits are infinite, but we will
return to this in the quantum case discussed below).

Given a set of classical solutions $ ( \p^{cl}(t), \x^{cl}(t) ) $,
and a phase space distribution function $w$, we are interested
in the probability that a classical solution will pass through
a region $\Delta $ of configuration space. We construct
this as follows. First of all we introduce the characteristic
function of the region $\Delta$,
$$
f_{\Delta} (\x )= \cases{1,&if $ \x \in \Delta $; \cr
0 &otherwise.}
\eqno(2.6)
$$
To see whether the classical trajectory $\x^{cl} (t)$
intersects this region, consider the phase space function
$$
A( \x, \p_0, \x_0) =
\int_{-\infty}^{\infty} dt \ \delta^{(n)} ( \x - \x^{cl} (t) )
\eqno(2.7)
$$
(In the case of periodic classical orbits, the range of $t$ is
taken to be equal to the period). This function
is positive for points $\x $ on the classical trajectory
labeled by $\p_0, \x_0 $
and zero otherwise.
Hence intersection of the classical
trajectory with the region $\Delta$ means,
$$
\int d^n \x \ f_{\Delta} (\x )
\int_{-\infty}^{\infty} dt \ \delta^{(n)} ( \x - \x^{cl} (t) ) > 0
\eqno(2.8)
$$
Or equivalently, that
$$
\int_{-\infty}^{\infty} dt
\ f_{\Delta} (\x^{cl} (t) ) > 0
\eqno(2.9)
$$
This quantity is essentially the amount of parameter time the
trajectory spends in the region $\Delta $.
We may now write down the probability for a classical
trajectory entering the region $\Delta $. It is,
$$
p_{\Delta} = \int d^n p_0 d^n x_0 \ w (\p_0, \x_0 )
\ \theta \left(
\int_{-\infty}^{\infty} dt
\ f_{\Delta} (\x^{cl} (t) ) - \e \right)
\eqno(2.10)
$$
In this construction, $\e$ is a small positive
number that is eventually sent to zero, and is included
to avoid possible ambiguities in the $\theta$-function
at zero argument. The $\theta$-function
ensures that the phase space integral is
over all initial data whose corresponding classical
trajectories spend a time greater than $\e$ in the
region $\Delta$.

The classical solution $ \x^{cl} (t) $ depends on
some fiducial initial coordinates and momenta,
$ \x_0 $ and $\p_0$, say. In the case of a
free particle, for example,
$$
\x^{cl} (t) = \x_0 + {\p_0 t \over M}
\eqno(2.11)
$$
The construction is independent of the choice of fiducial initial
points. If we shift $\x_0$, $\p_0$ along the classical
trajectories, the measure, phase space distribution function $w$
and the $\theta$-function are all invariant. Hence the integral over
$\x_0$, $\p_0$ is effectively a sum over classical trajectories.
The shift along the classical trajectories may also be thought of
as a reparametrization, and the quantity (2.10) is in fact a
reparametrization-invariant expression of the notion of a
classical trajectory. This means that the probability (2.10) has
the form of a phase space overlap of the ``state" with a
reparametrization-invariant operator.

It is useful also to write this result in a different
form, which will be more relevant to the results we get
in the quantum theory case. In the quantum theory,
we generally deal with propagation between fixed points
in configuration space, rather than with phase space point.
Therefore, in the free particle case, consider the change
of variables from $\x_0, \p_0$ to $\x_0, \x_f$, where
$$
\x_f = \x_0 + { \p_0 \over M} \tau
\eqno(2.12)
$$
Hence $\x_f$ is the position after evolution for
starting from $\x_0$ for parameter time $ \tau$.
The probability then becomes
$$
p_{\Delta} = {M \over \tau} \int d^n x_f d^n x_0 \ w (\p_0, \x_0 )
\ \theta \left(
\int_{-\infty}^{\infty} dt
\ f_{\Delta} (\x_0^f (t) ) - \e \right)
\eqno(2.13)
$$
where $\p_0 = M (\x_f - \x_0) / \tau $ and
$$
\x_0^f (t) = \x_0 + { (\x_f - \x_0) \over \tau} t
\eqno(2.14)
$$
The parameter $\tau$ may in fact be scaled out of the whole
expression, hence the probability is independent of it.

The result now has the form of an integral over ``initial'' and
``final'' points, analogous to similar results in quantum theory.
The result is again essentially a sum over classical trajectories
with the trajectories now labeled by any pair of points $\x_0$,
$\x_f$ along the trajectories, and is invariant under shifting
$\x_0$ or $\x_f$ along those trajectories. Naively, one might have
thought that the restriction to paths that pass through $\Delta$
is imposed by summing over all finite length classical paths
which intersect $\Delta $ as they go from the ``initial'' point
$\x_0$ to ``final'' point $\x_f$, that is, $\Delta$ lies {\it
between} the initial and final points. This is also what one might
naively expect in the quantum theory version. However, one can see
from the above construction that the correct answer is in fact to
sum over {\it all} classical paths (which can be of infinite
length) passing through $\x_0$ and $\x_f$ that intersect $\Delta$
at {\it any} point along the entire trajectory, even if $\Delta$
does not lie between the two points (see Fig.1). This feature is
related to the reparametrization invariance of the system.

The above point turns out to be quite crucial to what follows
in the rest of this paper, so it is worth saying it in an alternative
form. Loosely speaking, the statement is that only the entire
classical path respects the reparametrization invariance
associated with the constraint equation. A section of the
classical path does not. This may be expressed more precisely
in terms of the function $A(\x, \p_0, \x_0)$ introduced
in Eq.(2.7). This function is concentrated on the entire
classical trajectory, and is zero when $\x$ is not on the
trajectory.
It is easy to see that it has vanishing Poisson
bracket with the Hamiltonian $H = H (\p_0, \x_0)$,
since we have
$$
\eqalignno{
\{ H, A (\x, \p_0, \x_0) \}
& = \int_{-\infty}^{\infty} dt \ \{ H, \delta^{(n)} ( \x - \x^{cl} (t) ) \}
\cr
&= - \int_{-\infty}^{\infty} dt \ {d \over dt} \delta^{(n)} ( \x - \x^{cl} (t) )
\cr
& = 0
&(2.15) \cr}
$$
This is the precise sense in which the entire trajectory
is reparametrization invariant, and the phase space function
$A$ may be regarded as an observable -- a quantity which
commutes with the constraint $H$ [\cite{Rov3,Mar1}].
By way of comparison, consider a second phase space function
similarly defined, but on only a finite section of trajectory,
$$
B( \x, \p_0, \x_0) =
\int_{0}^{\tau} dt \ \delta^{(n)} ( \x - \x^{cl} (t) )
\eqno(2.16)
$$
It is easily seen that
$$
\{ H, B (\x, \p_0, \x_0) \}
= - \delta (\x - \x^{cl} (\tau) ) + \delta ( \x - \x^{cl} (0) )
\eqno(2.17)
$$
Hence $B$ ``almost'' commutes with $H$, failing only
at the end points, and it is in this sense that a finite
section of trajectory does not fully respect reparametrization
invariance.

A third version of the classical result is also useful. It is of
interest to obtain an expression for the probability for
intersecting an $(n-1)$-dimensional surface $\Sigma$. Since the
result (2.10) involves the parameter time spent in a finite volume
region $ \Delta $ it does not apply immediately. However, suppose
that the set of trajectories contained in the probability
distribution $w$ intersect the $(n-1)$-dimensional surface
$\Sigma$ only once. Then we may consider a finite volume region
$\Delta$ obtained by thickening $\Sigma$ along the direction of
the classical flow. If this thickening is by a small (positive)
parameter time $\Delta t$, then the quantity appearing in the
$\theta$-function in (2.10) is
$$
\eqalignno{
\int dt \int_{\Delta} d^n x \ \delta^{(n)} (\x - \x^{cl} (t))
&= \Delta t \int dt \int_{\Sigma} d^{n-1} x \ {\bf n} \cdot { d \x^{cl} (t)
\over dt } \ \delta^{(n)} (\x - \x^{cl} (t))
\cr & = \Delta t \ I [ \Sigma, \x^{cl} (t) ]
&(2.18) \cr}
$$
where ${\bf n}$ is the normal to $\Sigma$, and we suppose
that the normal is chosen so that ${\bf n} \cdot \ d \x^{cl} /dt $
is positive. The quantity
$I [\Sigma, \x^{cl} (t)] $, in a more general context,
is the intersection number of the
curve $\x^{cl} (t)$ with the surface $\Sigma$, and takes the value
$0$ for no intersections, or $\pm 1$ (depending on whether there is
an even or odd number of intersections). In this case we have
assumed that the trajectories intersect at most once, hence
$ I = 0 $ or $1$. We then have
$$
\theta ( \Delta t I  - \e ) = \theta ( I - \e' ) = I
\eqno(2.19)
$$
(where $\e = \Delta t \e'$)
and the probability for intersecting $\Sigma $ may be written
$$
p_{\Sigma} = \int dt \int d^n p_0 d^n x_0 \ w (\p_0, \x_0 ) \ \int_{\Sigma}
d^{n-1} x \ {\bf n} \cdot { d \x^{cl} (t)
\over dt } \ \delta^{(n)} (\x - \x^{cl} (t))
\eqno(2.20)
$$
At each $t$, we may perform a change of variables from $\p, \x$
to new variables $\p' = \p^{cl} (t)$, $\x' = \x^{cl} (t)$,
and using Eq.(2.5), we obtain the result
$$
p_{\Sigma} = {1 \over M} \int dt \int_{\Sigma} d^n p' \ d^{n-1} x' \ {\bf n}
\cdot \p' \ \ w (\p', \x' )
\eqno(2.21)
$$
Finally, the integrand is now in fact independent of $t$, so
the $t$ integral leads to an overall factor. (This might be infinite
but is regularized as discussed below). We therefore drop the
$t$ integral.

This result is relevant for the following reason. In the heuristic
``WKB interpretation'' of quantum cosmology, one considers WKB solutions
to the Wheeler-DeWitt equation of the form
$$
\Psi = C e^{iS}
\eqno(2.22)
$$
It is usually asserted that this corresponds to a set of classical
trajectories with momentum $ \p = \nabla S $, and with a probability
of intersecting a surface $\Sigma$ given in terms of the flux of the
wave function across the surface [\cite{Hal1,HaPa}].
As we shall show, from the decoherent
histories analysis, the quantum theory gives a probability for
crossing a surface $\Sigma$ proportional to Eq.(2.21) with $w$ replaced
by the Wigner function of the quantum theory. The Wigner function of
the WKB wave function is, approximately [\cite{Hal7}],
$$
W(\p, \x ) = \left | C ( \x ) \right|^2 \delta ( \p - \nabla S)
\eqno(2.23)
$$
Inserting in Eq.(2.21), we therefore obtain, up to overall factors,
the probability distribution,
$$
p_{\Sigma} =\int_{\Sigma} d^{n-1} x \ {\bf n} \cdot \nabla S
\ \left| C(\x) \right|^2
\eqno(2.24)
$$
We therefore have agreement with the usual heuristic analysis.

\head{\bf 3. The Quantum Case}

\subhead{\bf 3(A). Decoherent Histories Approach}

The decoherent histories approach to quantum theory is described
at length elsewhere [\cite{GeHa,Gri,Omn}], so only the briefest
review will be given here. The central object of interest is the
decoherence functional,
$$
D (\au, \au' ) =  {\rm Tr} \left( C_{ \au} \rho
C_{\au'}^{\dag} \right)
\eqno(3.1)
$$
where the histories are characterized by the class
operators $C_{\au}$, which satisfy
$$
\sum_{\au} C_{\au} = 1
\eqno(3.2)
$$
and therefore
$$
\sum_{\au, \au'} D (\au, \au') = {\rm Tr} \rho = 1
\eqno(3.3)
$$
In non-relativistic quantum mechanics, the class operators
are given by time-ordered sequences of
projection operators
$$
C_{\au} = P_{\a_n} (t_n) \cdots
P_{\a_1} (t_1)
\eqno(3.4)
$$
(and by sums of terms of this form), where
$\au$ denotes the string of alternatives $\a_1, \a_2 \cdots
\a_n$. The theory is however more general than this
and we will exploit this generality here.

Intuitively, the decoherence functional is a measure of the
interference between pairs of histories $\au$, $\au'$. When  its
real part is zero for $\au \ne \au' $, we say that the histories are
consistent and probabilities
$$
p (\au ) = D (\au, \au )
\eqno(3.5)
$$
obeying the usual probability sum rules may be assigned to them.
Typical physical mechanisms which produce this situation usually
cause both the real and imaginary part of $ D (\au, \au') $
to vanish. This condition is usually called decoherence
of histories, and is related to the existence of so-called
generalized records [\cite{GeHa,Hal6}].

In the non-relativistic case, for histories characterized by
projections onto configuration space, a path integral version of the
decoherence functional is available, and can be very useful. It has
the form,
$$
D( \au, \au') = \int_{\au} \D x \ \int_{\au'} \D y
\exp \left( \ih S[x(t)] - \ih S [y(t)] \right)
\rho (x_0, y_0)
\eqno(3.6)
$$
where the sum is over pairs of paths $x(t)$, $y(t)$ passing through
the pairs of regions $\au$, $\au'$. This is equivalent to the
form (3.1), (3.4), when the histories are strings of projections
onto ranges of positions. Eq.(3.6) is a useful starting point
for the generalization to timeless theories.

We would like to construct the decoherence functional
for the situation in question, in which we have a system
in an energy eigenstate, and we ask questions which do not
refer in any way to time. Two new issues arise in this case. The
first concerns the inner product through which the various piece of
the decoherence functional are put together. The second concerns
the construction of the class operators. We take each in turn.

\subhead{\bf 3(B). The Induced Inner Product}

For many situations, and especially
for the analogous situation in quantum cosmology, the
Hamiltonian has a continuous spectrum so the energy
eigenstates are not normalizable in the usual inner product,
$$
\la \Psi_1 | \Psi_2 \ra = \int d^n x \ \Psi_1^* (\x) \Psi_2 (\x )
\eqno(3.7)
$$
A way to deal with this has been developed, and goes by the name
of the induced inner product, or Rieffel induction
[\cite{Rie,HaMa}]. Consider the eigenvalue equation
$$
H | \Psi_{E \lambda} \ra = E | \Psi_{E \lambda} \ra
\eqno(3.8)
$$
where $\lambda$ denotes the degeneracy. These eigenstates
will typically satisfy
$$
\la \Psi_{E' \lambda'} | \Psi_{E \lambda} \ra =
\delta (E - E') \delta_{\lambda \lambda'}
\eqno(3.9)
$$
from which it is clear that the inner product diverges
when $E = E'$. The induced inner product on a set of
eigenstates of fixed $E$ is defined, loosely speaking,
by discarding the $\delta$-function $\delta (E-E')$.
That is, the induced or physical inner product is
then defined by
$$
\la \Psi_{E \lambda'} | \Psi_{E \lambda} \ra_{phys} =
\delta_{\lambda \lambda'}
\eqno(3.10)
$$
This procedure can be defined quite rigorously, and has been
discussed at some length in Refs.[\cite{Rie,HaMa}]. We will use it
here to construct the decoherence functional. A simple
prescription for using it in the decoherence functional is to
regularize each propagator and energy eigenstate by using a
different energy for each. The final answer will then involve a
number of $\delta$-functions in energy, as in (3.9), which are
simply dropped.

The induced inner product normalization for the wave functions
does in fact suggest a normalization scheme for the corresponding
classical phase space distribution function, which, recall, is not
normalizable in the case where the classical trajectories are
infinite (since $w$ is constant along those trajectories). The
idea is to consider the normalization of the Wigner function in
the quantum case. This is defined by
$$
W(\p,\X) = { 1 \over (2 \pi)^n } \int d^n v \ e^{- i\p \cdot \v} \
\rho( \X + \half \v, \X - \half \v)
\eqno(3.11)
$$
with inverse
$$
\rho(\x,\y) = \int d^np \ e^{ i \p \cdot (\x-\y) } \ W ( \p,
{\x+\y \over 2} )
\eqno(3.12)
$$
(See Refs.[\cite{BaJ,Tat}] for properties of the Wigner function.)
For an energy eigenstate $| \Psi_E \ra $ we first of all construct
a regularized density operator,
$$
\rho_{EE'} = | \Psi_E \ra \la \Psi_{E'} |
\eqno(3.13)
$$
which is normalized by
$$
{\rm Tr} \left( \rho_{E E'} \right) = \delta ( E - E')
\eqno(3.14)
$$
hence the corresponding Wigner function is normalized
according to
$$
\int d^n p d^n x \ W_{E E'} (\p, \x) = \delta (E - E')
\eqno(3.15)
$$
Now notice that the density operator obeys the equation
$$
[ H, \rho_{EE'} ] = (E - E') \rho_{E E'}
\eqno(3.16)
$$
Taking the Wigner transform of this equation we obtain
$$
{\cal L} W_{E E'} = i (E - E') W_{E E'}
\eqno(3.17)
$$
where ${\cal L}$ is the phase space operator
$$
{\cal L} =
\sum_k \left( - { p_k \over M} { \partial \over \partial x_k}
+ { \partial V \over \partial x_k} { \partial \over \partial p_k}
\right) + {\cal L}_q
\eqno(3.18)
$$
It is a sum of the classical Liouville operator, plus a term
${\cal L}_q $ describing quantum modifications.

We may now see how to normalize the classical case. We take the
classical distribution function to be described by Eq.(3.17)
with the quantum term ${\cal L}_q$ set to zero. The Liouville
operator may then be written,
$$
{\cal L} = - { d \over ds}
\eqno(3.19)
$$
for some parameter $s$, and Eq.(3.17) may be solved, with the
result,
$$
w_{E E'} = e^{ - i s(E - E') } w_{EE}
\eqno(3.20)
$$
The exponential factor now effectively regularizes the phase
space distribution function. $w_{EE}$ is constant along the
classical trajectories, but $w_{E E'}$ is not, and, in the
normalization (3.15), the part of the integral along the trajectories
is an integral over $s$ which produces the $\delta$-function.

\subhead{\bf 3(C). Construction of the Decoherence Functional.}

The next part of the construction of the decoherence functional is
the class operators, $C_{\a}$. These are to describe histories of
fixed energy which do or do not pass through the region $\Delta$
without regard to time, denoted $ \a = \Delta $ and $\a = \bar
\Delta$ respectively. We follow Refs.[\cite{Har3,HaMa}], in part.

Consider
first the amplitude to go from $\x_0$ at time $t=0$
to $\x_f$ at time $t= \tau$ passing through the
region $\Delta$, or not, at any time in between. This
is given by
$$
g_\a ( \x_f, \tau | \x_0, 0 )
= \int_\a \D x (t) \exp \left( \ih S [x(t)] \right)
\eqno(3.21)
$$
where the sum is over all paths $x(t)$ with $t$ in the
range $[0, \tau]$ which pass through $\Delta$,
or never pass through $\Delta$. It therefore
satisfies,
$$
\sum_{\a} g_{\a} = g_{\Delta} + g_{\bar \Delta} = g
\eqno(3.22)
$$
where $ g = g ( \x_f, \tau | \x_0, 0 ) $ is the unrestricted
propagator. There
are many ways of constructing this sort of object
more explicitly (see Refs.[\cite{HaTh,HaOr,AuKi}] for example), but
here it is useful to exploit the construction used in
the classical case. The amplitude to pass through
$\Delta $ is therefore given by
$$
g_\Delta ( \x_f, \tau | \x_0, 0 )
= \int \D x (t) \exp \left( i S [x(t)] \right)
\ \theta \left(
\int_0^{\tau} dt
\ f_{\Delta} (\x (t) ) - \e \right)
\eqno(3.23)
$$
Here, the $\theta$-function ensures that only paths
$x(t)$ that spend a time in excess of $\e$ in
$\Delta$ contribute to the sum.

The class operator $ C_{\au}$ is a propagator at fixed energy, $E$,
say, so this is given by
$$
\la \x_f | C_{\au} | \x_0 \ra
=
\int_{-\infty}^{\infty} d \tau
\ e^{ - i E \tau } \ g_{\a} ( \x_f, \tau | \x_0, 0 )
\eqno(3.24)
$$
When $g_{\a}$ is replaced with an unrestricted propagator, we
require that (3.24) is annihilated by $H-E$, and this is why we
choose an infinite range for $\tau$, rather than a half-infinite
one [\cite{Hal3,Tei,HalH}]. (As we shall see below, when $g_{\a}$
is a restricted propagator, we encounter some difficulties here,
although the correct range for $\tau$ is still the infinite one).
The total (regularized) decoherence functional is therefore given
by
$$
\eqalignno{
D (\a, \a') = & \int d^n x_f d^n x_0 d^n x_0'
\int_{-\infty}^{\infty} d \tau
\int_{-\infty}^{\infty} d \tau'
\ e^{ - i E \tau } \ e^{  i E' \tau' }
\cr
\times
& \ g_{\a} ( \x_f, \tau | \x_0, 0 )
\ g^*_{\a'} ( \x_f, \tau' | \x_0', 0 )
\ \Psi_{E_0} (\x_0) \Psi^*_{E_0'} (\x_0')
&(3.25) \cr}
$$
This may also be written,
$$
\eqalignno{
D (\a, \a') = &
\int_{-\infty}^{\infty} d \tau
\int_{-\infty}^{\infty} d \tau'
\ e^{ - i E \tau } \ e^{  i E' \tau' }
\cr
\times
&
\int_{\a} \D \x(t) \ \int_{\a'} \D \x'(t)
\exp \left( i S_0^\tau [\x(t)] - i S_0^{\tau'}
[\x'(t)] \right)
\cr \times
&
\ \Psi_{E_0} (\x_0) \Psi^*_{E_0'} (\x_0')
&(3.26) \cr}
$$
where note that the two actions in the path integral
are over different ranges of time.
It is straightforward to show that
$$
\sum_{\a, \a'} D (\a, \a') =
\delta (E-E_0) \delta (E'-E_0') \delta (E_0 - E_0')
\eqno(3.27)
$$
In the induced inner product scheme we therefore
replace the righthand side by $1$, verifying that the construction
is correctly normalized.

\subhead{\bf 3(D). Modified Class Operators}

The basic scheme described above runs into a interesting
difficulty in that the class operators defined by (3.24) do not
satisfy the constraint equation.
We have, for example, for the class operator for
paths that enter the region $\Delta$,
$$
C_{\Delta} (\x_f, \x_0) =
\int_{-\infty}^{\infty}
d \tau \ e^{- i E \tau} \ \int \D x (t) \exp \left( i S [x(t)] \right)
\ \theta \left(
\int_0^{\tau} dt
\ f_{\Delta} (\x (t) ) - \e \right)
\eqno(3.28)
$$
It may be shown that this satisfies the constraint everywhere
except on the boundaries of the region $\Delta$. That is, there is a
discontinuity as one of the end-points crosses the boundary of $\Delta$.
This is an issue because a basic rule of the game of constructing
the decoherence functional for these systems is that we only work with objects
which satisfy the constraint equation (or operators which commute with it).
Mathematically, this is to ensure that
the underlying symmetry, reparametrization invariance, is fully respected
(the induced inner product, note, is defined only between objects which
satisfy the constraint).
Physically, it is related to the fact that the universe is a closed system,
and measurements of it (the class operators are generalizations of the
notions of measurement) must not displace its wave function.

Because of this difficulty, it is necessary to define a modified
class operator $C_{\Delta}'$ which is as close as possible to
the path integral one, but satisfies the constraint equation everywhere
[\cite{HaMa}].
A way to do this is to first compute the class operator
$C_{\Delta}$ when one of the end-points is inside $\Delta$. This
defines a solution to the constraints for one end-point inside and
the other outside. We then {\it define} $C_{\Delta}'$ to be the
object which satisfies the constraint everywhere and matches
this expression for one end-point inside and the other
outside.

An example of this difficulty was encountered and solved in
Ref.[\cite{HaTh}], which concerned the decoherent histories analysis of the
relativistic particle. Suppose one considers the class operator
for propagating between spacetime points $x^{\mu}_0$ and $x^{\mu}_f$,
with the
restriction that the paths never cross the spacelike surface $x^0
= constant$. This object, denoted $C_r (x_f, x_0)$, is readily
constructed using the free propagator together with the method of
images. However, it does not satisfy the constraint everywhere:
the constraint operating on $C_r$ gives a $\delta$-function on the
spacelike surface. One can also see physically why there might be
problems here. Suppose for simplicity that $x_f $ and $x_0$ are
timelike separated and consider the trajectory of a classical
particle passing through these two points. Such a trajectory, if
extended beyond these points, must cross {\it every} surface of
constant $x^0$. Similarly, in
the quantum theory, there are no non-trivial solutions to the
Klein-Gordon equation that are zero on one side of a spacelike
surface. Hence the only sensible possible answer for the class
operator for this situation is $C_r = 0$. This is indeed the
solution used in Ref.[\cite{HaTh}] and led to physically
expected results.

This issue concerning the replacement of the original class operators
with modified ones is related to the fact that in the original
class operator (3.28), the functions restricting the paths to enter
the region $\Delta$ involve a time integral over a finite interval
$[0, \tau]$, whereas the classical result (2.13) involves a similar
restricting function but with a time integral over an infinite range.
If a finite range is used in the classical result, it is no longer
reparametrization invariant, as we saw in Section 2. This is why
we discussed the classical result at such length.

The construction of these modified class operators is crucial to the
construction of the decoherence functional for reparametrization-invariant
theories. It is probably fair to say that the method suggested above
for constructing them
has not at this stage been fully explored. We will show below that
a physically plausible modified class operator is readily constructed
in the semiclassical approximation, but a more thorough investigation of this
issue will be deferred to a later publication. It should also be noted
that reparametrization-invariance can be rather subtle. For
example, the path-integral constructed class operators
Eq.(3.28) certainly {\it appear} to be reparametrization-invariant
at the level of symmetry transformations at the Lagrangian
level, yet do not quite satisfy the constraint. Here, we have
used the expression ``reparametrization-invariant'' to mean
satisfying the constraint everywhere (or having zero Poisson
bracket with the Hamiltonian everywhere, in the classical case).
This issue is related to the connections between Lagrangian and
Hamiltonian symmetries, and between the path integral and
Dirac quantizations. See Refs.[\cite{Har3,HalH}] for further discussion.

\head{\bf 4. The Semiclassical Approximation to the
Decoherence Functional and Probabilities}

We may now compute the decoherence functional. It is
$$
D(\a, \a') = \int d^n x_f d^n x_0 d^n y_0
\ C_{\a}' (\x_f,\x_0 ) C_{\a'}'^* (\x_f, \y_0)
\ \Psi (\x_0) \Psi^* (\y_0)
\eqno(4.1)
$$
(For convenience, here and in what follows we will in fact
drop the notation involving different values of $E$ to
regularize the expressions, unless necessary).
To see how the semiclassical approximation works out, we will
in this Section
{\it assume} decoherence (for example, by restricting to
special initial states) and concentrate on
the construction of the probabilities that the system passes
through the region $\Delta$. We will return to the question of
decoherence for general initial states in Section 6.

We begin by computing the semiclassical approximation to the modified
class operator $C_{\Delta}'(\x_f,\x_0)$. Recall that it is given by
a suitable modification of the
path integral expression (3.28) with one end-point inside $\Delta$
and the other outside. In the absence of any restrictions on the
paths, the path integral will be dominated by the classical
paths connecting the initial and final points. The classical
paths will be the solutions to the equations of motion
$$
M \ddot \x + \nabla V(\x) = 0
\eqno(4.2)
$$
which satisfy the boundary conditions
$$
\x ( 0 ) = \x_0, \quad \x ( \tau ) = \x_f
\eqno(4.3)
$$
In addition, these paths must satisfy the constraint
equation
$$
\half M \dot \x^2 + V ( \x ) = E
\eqno(4.4)
$$
This equation determines the time $ \tau $ in terms
of $\x_0$, $\x_f$ and $E$, hence the final form of
the extremizing paths have no reference to time.
It is also useful to introduce
$ A(\x_f, \x_0)$, the classical action
from $\x_0$ to $\x_f$. It obeys the time-independent Hamilton-Jacobi
equation with respect to each end point,
$$
{1 \over 2M} \left( \nabla A \right)^2 + V = E
\eqno(4.5)
$$
The initial and final momenta are given by derivatives of the
classical action,
$$
\p_f = \nabla_f A (\x_f, \x_0), \quad \p_0 = - \nabla_0 A (\x_f, \x_0)
\eqno(4.6)
$$

The semiclassical approximation to the unrestricted path integral
is given by a sum of terms each of the form
$$
G (\x_f, \x_0) = P(\x_f, \x_0) \ e^{ i A(\x_f, \x_0)}
\eqno(4.7)
$$
The quantity $P$ is a prefactor, whose specific form is lengthy
to calculate, but will not in fact be required.
(For the case of the time-dependent propagator, the prefactor
would be given in terms of the determinant of the matrix of
second derivatives of $ A(\x_f, \x_0)$. Here, because of the
constraint equation (4.5), this matrix is in fact singular
and the expression for the prefactor is more complicated
[\cite{Bar,Sch}]).

The semiclassical form (4.7) satisfies the constraint
equation ({\it i.e.}, is annihilated by $H-E$)
in the WKB approximation, as it should.
If the classical trajectory from $\x_0$ to $\x_f$
is unique there will be just one term in the semiclassical
approximation. If there is more than one, there will be a sum
of similar terms, one for each trajectory. For the moment
we will assume that there is just one. We will also
assume that the extremizing classical solution is real, and thus the
action $A(\x_f, \x_0) $ is real.

With the restriction that the paths must pass through $\Delta$,
we expect that the class operator will be given again by (4.7)
when the classical path passes through $\Delta$, and will
be zero when the classical path does not pass through $\Delta$.
It is then not difficult to see that the
modified class operator for this case may therefore be written,
$$
C_{\Delta}' (\x_f, \x_0) = \theta \left( \int_{-\infty}^{\infty} dt
\ f_{\Delta} (\x_0^f (t) ) -\e \right)
\ P(\x_f, \x_0) \ e^{ i A(\x_f, \x_0)}
\eqno(4.8)
$$
The $\theta$-function here is the same as in the (rewritten) classical
case Eq.(2.13) in terms of ``initial'' and ``final'' points, where
$\x_0^f (t)$ denotes that classical
path from $\x_0$ to $\x_f$. (This is exactly as in the classical
case depicted in Fig.1). Note also that
$$
\nabla A \cdot \nabla
\ \ \theta \left( \int_{-\infty}^{\infty} dt
\ f_{\Delta} (\x_0^f (t) ) -\e \right) = 0
\eqno(4.9)
$$
as may be shown by shifting the $t$ integration.
It follows that the modified class operator is a WKB solution
to the constraint equation, as required. We have therefore
succeeded in computing, in the semiclassical approximation,
a modified class operator satisfying the constraint equation
everywhere, corresponding to the restriction to paths
passing through the region $\Delta$. This is a very simple
result but turns out to be crucial to the rest of the derivation.

It is important that
$t$ is integrated over an infinite range in the quantity inside
the $\theta$-function, otherwise the modified class operator would not
in fact satisfy the constraint. Recall that the originally
defined class operator Eq.(3.24) contained a similar $\theta$-function,
with a finite range of time integration, which one might have been
tempted to use in the semiclassical approximation, but this class operator
does not in fact satisfy the constraint.

Hence we see that the
difference between the modified and original class operators
in the semiclassical approximation is the difference between using
the entire classical trajectory or using finite segments of it
in the $\theta$-functions. We also see that these modified
class operators are the correct ones to use in order to
be consistent with the discussion of the classical case
and Eq.(2.13). There, we saw that it is appropriate to
sum over classical paths intersecting $\Delta $
even if $\Delta$ does not lie
on the segment of classical trajectory {\it between}
$\x_0$ and $\x_f$.
This feature therefore appears to be
necessary for the particular type of
reparametrization invariance used here.
Only the entire trajectory is reparametrization-invariant
notion. A finite section of trajectory is not.
(See Ref.[\cite{HalH}] for a further discussion of
reparametrizations in this sort of context).

The off-diagonal terms of the
decoherence functional are now given in the semiclassical
approximation by
$$
\eqalignno{
D( \Delta, \bar \Delta )  &= \int d^n x_f d^n x_0 d^n y_0
\ \theta \left( \int_{-\infty}^{\infty}
dt \ f_{\Delta} (\x_0^f (t) ) -\e \right)
\cr
& \times \left( 1 - \theta \left( \int_{-\infty}^{\infty}
dt \ f_{\Delta} (\y_0^f (t) ) -\e \right) \right)
\cr
& \times
P(\x_f, \x_0) P^* (\x_f, \y_0)
\ \exp \left( i A (\x_f, \x_0) - i A(\x_f, \y_0) \right)
\ \rho( \x_0, \y_0)
&(4.10) \cr }
$$
It is now essentially a sum over pairs of classical paths, and the
$\theta$-functions restrict the paths to either pass on not
pass through the region $\Delta$. We see also from the invariance
property of the $\theta$-functions, (4.9), that they
are invariant under shifting the regions $\Delta_{\a}$ along their
classical trajectories. This is the expression of the idea that
the decoherence functional, in the semiclassical approximation,
knows only about entire trajectories.

It is now convenient to introduce the variables
$$
\X_0 = \half ( \x_0 + \y_0), \quad \v = \x_0 - \y_0
\eqno(4.11)
$$
and thus
$$
\x_0 = \X_0 + \half \v, \quad \y_0 = \X_0 - \half \v
\eqno(4.12)
$$
and we also rewrite the density operator in terms of the Wigner function
Eq.(3.11).
We will discuss the detailed mechanism of decoherence in the next
Section. For the moment, we will simply assume decoherence,
which essentially means assuming that $\v$ is concentrated around zero,
and work
out the form of the probabilities. Although we note that this assumption
can be justified for initial states $\rho ( \x_0, \y_0)$ which are
approximately diagonal in position.
We now set $\v = 0 $ in
the prefactors $P$ and in the $\theta$-functions, and
we obtain the probability
$$
\eqalignno{
p_{\Delta} &= \int d^n x_f d^n X_0 d^n v d^n p
\ \theta \left( \int_{-\infty}^{\infty}
dt \ f_{\Delta} (\X_0^f (t) ) -\e \right)
\left| P(\x_f, \X_0) \right|^2
\cr & \times
\ \exp \left( i A (\x_f, \X_0 + \half \v) - i A(\x_f, \X_0 - \half \v)
+ i \p \cdot \v \right)
\ W (\p, \X_0)
&(4.13) \cr }
$$
Expanding the action terms to linear order in $\v$ (decoherence
again allows us to drop the higher order terms), the $\v$ integral
may be performed and we obtain
$$
\eqalignno{
p_{\Delta} &= \int d^n x_f d^n X_0 d^n p
\ \theta \left( \int_{-\infty}^{\infty}
dt \ f_{\Delta} (\X_0^f (t) ) -\e \right)
\cr & \times
\left| P(\x_f, \X_0) \right|^2
\delta^{(n)} \left( \p + \nabla_0 A (\x_f, \X_0) \right)
\ W (\p, \X_0)
&(4.14) \cr }
$$
where $\nabla_0$ operates on the initial point $\X_0$.
Finally, the integration over $\x_f$ may be performed.
The delta-function constraint then means that the
quantity $\X_0^f (t)$ (the classical path from $\X_0$
to $\x_f$) is replaced by $\X^{cl} (t)$ (the classical
path with initial data $\X_0, \p_0$). Although we have
not worked out the explicit form of the prefactor $P$,
we deduce that it must in fact drop out when
the $\x_f$ integration is carried out, because
the probability must equal $1$ when the $\theta$-function
is removed. (From this we deduce that $| P (\x_f, \X_0)|^2$
must be the Jacobian factor in the change of integration
variables from $\x_f$ to $\nabla_0 A (\x_f, \X_0)$).
We therefore obtain the final
result,
$$
p_{\Delta} = \int d^n X_0 d^n p
\ \theta \left( \int_{-\infty}^{\infty}
dt \ f_{\Delta} (\X^{cl} (t) ) -\e \right)
\ W (\p, \X_0)
\eqno(4.15)
$$
As expected, this is the classical result Eq.(2.10) with the
classical phase space distribution function replaced
by the Wigner function.

% [ROLE OF RIEFFEL INDUCTION]

\head{\bf 5. Systems of Harmonic Oscillators}

There is one simple system for which the discussion of decoherence
and probabilities is particularly simple, and this is the case
of a collection of harmonic oscillators. It also enjoys the
property that its spectrum is discrete, hence the induced inner
product is not required for normalization.

In non-relativistic quantum mechanics, in the search for emergent
quasi-classical histories, it is of interest to consider histories
characterized by strings of phase space quasi-projectors $P_{\Gamma}$.
These are positive hermitian operators concentrated on
a region $\Gamma$ of phase space, but are not quite projectors
since they only have $P^2_{\Gamma} \approx P_{\Gamma}$. Omn\`es
has proved an important theorem about these projectors, which is
essentially that they are approximately preserved in form under
unitary evolution and moreover approximately
follow classical evolution [\cite{Omn}]. That is,
$$
e^{ - i H t } P_{\Gamma} e^{ i H t } \approx P_{\Gamma_{t}}
\eqno(5.1)
$$
where $\Gamma_{t}$ is the classical evolution of the phase
space cell $\Gamma$. The approximation holds when the phase
space cells are significantly larger than a quantum sized cell,
and for times not so long that wave packet spreading becomes
significant. In the special case of the harmonic oscillator,
the approximation holds for all time. This result allows
one to show that firstly, histories of phase space projectors
are approximately decoherent for a wide variety of initial states,
and secondly, that their probabilities are peaked about classical
evolution. Differently put, on sufficiently coarse grained
scales, quantum systems have an approximate determinism
that ensures decoherence and approximate correspondence with
classical physics.

It seems reasonable to suppose that the timeless models considered
here might have analogous properties. We will demonstrate this
for a system of harmonic oscillators in an energy eigenstate.
The Hamiltonian for a set of $N$ identical harmonic oscillators is
$$
H_0 = \half \left( \p^2 + \x^2 \right) - \half N
\eqno(5.2)
$$
(the factor of $\half N$ is included to subtract the vacuum
state energy and avoid certain phase factors).
The standard coherent states (see Ref.[\cite{Gar}], for example)
are denoted $ | \p, \x \ra $ and they
have the important property that they are preserved in form under
unitary evolution,
$$
e^{ - i H_0 t } | \p, \x \ra = | \p_t, \x_t \ra
\eqno(5.3)
$$
where $ \p_t, \x_t $ are the classical solutions matching
$\p, \x $ at $t=0$, hence they are strongly peaked about
the classical path. In Ref.[\cite{Hal2}], a set of states were introduced
which are timeless analogues of the usual coherent states.
They are
$$
\eqalignno{
| \phi_{\p \x} \ra &=  \delta (H_0 - E ) | \p, \x \ra
\cr
&=  \int_0^{2 \pi} { dt \over 2 \pi } \ e^{ - i (H_0 - E ) t }
\ | \p , \x \ra
\cr
&=   \int_0^{2 \pi} { dt \over 2 \pi } \ e^{i E t }\ | \p_t, \x_t \ra
&(5.4) \cr}
$$
These states were referred to in Ref.[\cite{Hal2}] as ``timeless coherent
states''. (See also [\cite{Rov1,Kla}]).
They are exact eigenstates of the Hamiltonian,
$$
H_0 | \phi_{\p \x} \ra = E | \phi_{\p \x} \ra
\eqno(5.5)
$$
Furthermore, since the coherent states $ | \p_t, \x_t \ra $ are
concentrated at a phase space point for each $t$, integrating $t$
over a whole period produces a state which is concentrated along
the entire classical trajectory. They are therefore the natural
analogues of the usual coherent states. Their properties are
similar in many ways to the usual coherent states and are
described in more detail in Ref.[\cite{Hal2}].

Each state is labeled by a fiducial phase
space point $\p, \x$ which
determines the classical trajectory the state is peaked about.
Under evolution of the fiducial point $\p, \x$ to another point,
$\p_s, \x_s$, say, along the same classical trajectory, the
state changes by a phase,
$$
| \phi_{\p \x} \ra \ \ria \  | \phi_{\p_s \x_s} \ra =
e^{ i E s } | \phi_{\p \x} \ra
\eqno(5.6)
$$
Two timeless coherent states of different energy are exactly
orthogonal. If they have the same energy
then they are approximately
orthogonal if they correspond to sufficiently distinct classical
solutions. They also obey a completeness relation,
$$
\int {d^N p \ d^N x \over (2 \pi )^N }
\  |\phi_{\p\x} \ra \la \phi_{\p \x} |  =  \delta (H_0 - E)
\eqno(5.7)
$$
[Note that the notation $ \delta (H_0 -E)$ is a rather loose one.
This object is really the projection operator onto the subspace
of energy $E$, for which it is exactly true that
$ \delta (H_0-E)^2 = \delta (H_0 - E)$].
Since $ \delta (H_0 - E ) | \psi \ra = | \psi \ra $ on any solution
to the eigenvalue equation $(H_0 - E) | \psi \ra = 0 $, this is
essentially a completeness relation on the set of solutions
to the eigenvalue equation.
We may therefore write any solution
$|\psi \ra$ as a superposition of timeless coherent states,
$$
| \psi \ra = \int {d^N p \ d^N x \over (2 \pi )^N}
\  | \phi_{\p \x } \ra \la \phi_{\p \x} |\psi \ra
\eqno(5.8)
$$

Given these preliminaries, we may now discuss the decoherence
functional. We will consider coarse grainings in which the
paths in configuration space either pass or do not pass through
a series of regions denoted $\Delta = \Delta_1, \Delta_2
\cdots $, we will take $\Delta_1, \Delta_2 \cdots $ to lie
along a classical path. Hence we need at least two such regions
to fix a configuration space path.

The decoherence functional, in terms of the modified
class operators, is, in the semiclassical approximation,
$$
D(\Delta, \bar \Delta) = \int d^n x_f d^n x_0 d^n y_0
\ C_{\Delta}' (\x_f,\x_0 ) C_{\bar \Delta}'^* (\x_f, \y_0)
\ \Psi (\x_0) \Psi^* (\y_0)
\eqno(5.9)
$$
The modified class operator $C'_{\Delta}(\x_f, \x_0) $ is given
by Eq.(4.8),
so is equal to the unrestricted semiclassical propagator $G$
when $\x_f$ or $\x_0$ lie on the
classical path specified by $\Delta$ and is zero otherwise.
Also, $ C_{\bar \Delta}' = \delta (H_0 - E) - C_{\Delta}'$.

We first consider the case in which the initial state is
a timeless coherent state, $| \phi_{\p \x} \ra $.
It is then straightforward to see that
$$
C_{\Delta}' | \phi_{\p \x} \ra \approx | \phi_{\p \x} \ra
\eqno(5.10)
$$
when the trajectory labeled by the fiducial points $\p, \x$
passes through the regions $\Delta$, and
$$
C_{\Delta}' | \phi_{\p \x} \ra \approx 0
\eqno(5.11)
$$
otherwise. Also, since
$$
\delta (H_0 - E) | \phi_{\p \x} \ra = | \phi_{\p \x} \ra
\eqno(5.12)
$$
it follows that
$$
C_{\bar \Delta}' | \phi_{\p \x } \ra \approx 0
\eqno(5.13)
$$
when the trajectory labeled by $\p, \x $ passes through
$\Delta$. From these results it is easy to see that
the decoherence functional is approximately diagonal.
Furthermore, the probability for entering the regions
$\Delta$ is then approximately $1$ or $0$, depending on whether
the classical trajectory of the timeless coherent state
passes through $\Delta$.

Now consider the case of a more general initial state. We expand
it in timeless coherent states, as in Eq.(5.8). Using the above
results, we therefore find
$$
C_{\Delta}' | \psi \ra  \approx
\int_{D} {d^N p \ d^N x \over (2 \pi )^N}
\  | \phi_{\p \x } \ra \la \phi_{\p \x} |\psi \ra
\eqno(5.14)
$$
Here, $ D $ denotes the set of phase space
points $\p, \x$ whose classical trajectories pass through the
regions $\Delta$ in configuration space. Similarly,
$$
C_{\bar \Delta}' | \psi \ra  \approx
\int_{\bar D} {d^N p \ d^N x \over (2 \pi )^N}
\  | \phi_{\p \x } \ra \la \phi_{\p \x} |\psi \ra
\eqno(5.15)
$$
where $\bar D$ denotes the set of phase space points whose
classical trajectories never pass through $\Delta$.
Clearly we again have approximate decoherence, because of
the approximate determinism. We may therefore assign
a probability for passing through $\Delta$,
$$
p_{\Delta} \approx \int_{D} {d^N p \ d^N x \over (2 \pi )^N}
\left| \la \phi_{\p \x} |\psi \ra \right|^2
\eqno(5.16)
$$
It is the integral over the phase space region $\Delta$
of the phase space distribution function
$ \left| \la \phi_{\p \x} |\psi \ra \right|^2$.
Because $| \psi \ra $ is an eigenstate of the
Hamiltonian, it is easy to see
using the definition (5.4) of the timeless coherent
states that
$$
\la \phi_{\p \x} | \psi \ra = \la \p \x | \psi \ra
\eqno(5.17)
$$
and so the probability now is
$$
p_{\Delta} \approx \int_{D} {d^N p \ d^N x \over (2 \pi )^N}
\left| \la \p \x |\psi \ra \right|^2
\eqno(5.18)
$$
It is then a standard result that the integrand is in fact a smeared
Wigner function
$$
\left| \la \p \x |\psi \ra \right|^2
= \int d^N p' d^N x' \ e^{ - \half ( \p -\p')^2 - \half (\x - \x')^2
} \ W (\p', \x')
\eqno(5.19)
$$
This object is positive even though the original Wigner function
$W$  of $| \psi \ra $ is not [\cite{Gar}].
Hence we obtain a result which is essentially identical
to the classically anticipated result (2.10), with a smeared
Wigner function as the phase space distribution function.

It is also of interest to note that the result for the probability
may be written in the form
$$
p_{\Delta} = \la \psi | P_{D} | \psi \ra
\eqno(5.20)
$$
where $P_{D}$ is the approximate projection operator
$$
P_D = \int_D {d^N \p \ d^N \x \over (2 \pi )^N }
\  |\phi_{\p\x} \ra \la \phi_{\p \x} |
\eqno(5.21)
$$
Moreover, since the timeless coherent states $ | \phi_{\p \x} \ra$
are exact eigenstates of $H_0$, we have that
$$
[ P_D, H_0 ] = 0
\eqno(5.22)
$$
Hence, we see that the result may be written in the standard
quantum-mechanical form for a probability, in terms of
an operator which commutes with the constraint.

The result here, of approximate decoherence and simple
expressions for the probabilities, like the corresponding
non-relativistic result is due to the approximate determinism
contained in the quantum theory. It works only when we
ask for the probabilities for approximately classical
histories. To obtain probabilities for more complicated
histories, and for systems which are not harmonic
oscillators (where there is wave packet spreading),
we need an environment to produce decoherence.

% [FIX NORMALIZATION OF TIMELESS COHERENT STATES. GENERAL PROBABILITIES]

\head{\bf 6. Decoherence Through an Environment}

As stated above, the decoherence functional is typically
not diagonal for most initial states, and a physical mechanism is required
to produced decoherence. In this section we therefore
consider the addition of an environment to produce decoherence of histories.
The results of this Section therefore simply justify the assumed
decoherence of Section 4, and little affect the final result of the
probabilities, but it is important to see in detail how this
works.

\subhead{\bf 6(A). Semiclassical Approximation to the Decoherence
Functional with Environment}

For what we will do here, the specific form of the environment
turns out not to be very important. But for definiteness,
we take the environment to be a large collection of harmonic oscillators
with coordinates denoted $\q_A$, where $A$ runs over a large number
of values,
with a linear coupling to the system. For notational simplicity
we will assume that for each system coordinate
$\x$ in the $n$-dimensional configuration space
there is a set of $n$ oscillators with
coordinate $\q$ for the environment. The case of
of more oscillators is easily obtained from this.
The total action of the
system is
$$
S = S_0 [\x ] + S_{_\E}[\x,\q]
\eqno(6.1)
$$
and the corresponding Hamiltonian is
$$
H = H_0 (\x)  + H_{_\E} (\x, \q)
\eqno(6.2)
$$
We shall assume that the state of the whole system has the form
$$
\Psi (\x, \q) = \psi (\x) \chi (\x, \q)
\eqno(6.3)
$$
This may be inserted into the Wheeler-DeWitt equation $(H-E) \Psi = 0$
to obtain a perturbative solution about the solution with no environment.
We will concentrate on the case in which the wave
function is of oscillatory form, so the background solution is
of WKB form
$$
\psi (\x) = C(\x ) e^{ i S (\x) }
\eqno(6.4)
$$
where $S$ obeys the Hamilton-Jacobi equation
$$
\half \left( \nabla S \right)^2 + V(\x) = E
\eqno(6.5)
$$
and $C$ obeys the equation
$$
\nabla^2 C + 2 \nabla S \cdot \nabla C = 0
\eqno(6.6)
$$
The environment wave functions obey the Schr\"odinger equation
$$
i \nabla S \cdot \nabla \chi = H_{_\E}\  \chi
\eqno(6.7)
$$
We will consider the case of a superposition of WKB states in Section 7.

The decoherence functional now is
$$
\eqalignno{
D (\a, \a') = &
\int_{-\infty}^{\infty} d \tau
\int_{-\infty}^{\infty} d \tau'
\ e^{ - i E \tau } \ e^{  i E' \tau' }
\cr & \times
\int_{\a} \D \x(t) \ \int_{\a'} \D \y(t)
\exp \left( i S_0^\tau [\x(t)] - i S_0^{\tau'}
[\y(t)] \right)
\cr & \times
\ F [ \x(t), \y(t), \tau, \tau' ]
\ \psi_{E_0} (\x_0) \psi^*_{E_0'} (\y_0)
&(6.8) \cr}
$$
(suspending for the moment the necessity to use modified
class operators). Here $S_0^{\tau}$ denotes the action over the fixed time
range $[0, \tau]$ (and note that the time ranges are different
on either side of the decoherence functional).
The influence functional $F[ \x(t), \y(t), \tau, \tau' ]$ is given by,
$$
\eqalignno{
F[ \x(t), \y(t), \tau, \tau' ] = & \int \D \q(t) \D \r (t)
\exp \left( i S_{\E}^{\tau} [\x, \q] - i S_{\E}^{\tau'} [\y, \r]
\right)
\cr &
\quad \quad \times \ \chi (\x_0, \q_0 ) \chi^* (\y_0, \r_0)
&(6.9) \cr}
$$
It is different in form in two ways to the usual influence
functional. Firstly, the time ranges on either side are not the
same, and secondly, the initial state $\chi$ depends on the
system variables $\x$, with a different dependence on either side
of the influence functional. The functional integral is over all
pairs of paths $\q (t), \r(t) $ which meet at the final point
$$
\q (\tau) = \r(\tau')
\eqno(6.10)
$$
and this point is summed over. The paths also
match the initial values
$\q_0, \r_0$, which are then folded into the initial
state.

It is useful to go now to the semiclassical approximation for
the system variables. We also now recall that we must use modified
class operators for the system variables (this does not affect
the environment dynamics, at this level of approximation).
The off-diagonal terms of the decoherence functional are now given by
$$
\eqalignno{
D (\Delta, \bar \Delta)  &= \int d^n x_f d^n x_0 d^n y_0
\ \theta \left( \int_{-\infty}^{\infty}
dt \ f_{\Delta} (\x_0^f (t) ) -\e \right)
\cr
& \times \left( 1 - \ \theta \left( \int_{-\infty}^{\infty}
dt \ f_{\Delta} (\y_0^f (t) ) -\e \right) \right)
\cr
& \times
P(\x_f, \x_0) P^* (\x_f, \y_0)
\ \exp \left( i A (\x_f, \x_0) - i A(\x_f, \y_0) \right)
\cr
& \times
\ F (\x_f, \x_0, \y_0 )
\ \psi ( \x_0) \psi^* ( \y_0)
&(6.11) \cr }
$$
Here, $F (\x_f, \x_0, \y_0 )$ is the influence functional with
the semiclassical approximation for the system variables inserted.
That is, for $\x(t)$ we insert the classical trajectory from
$\x_0$ to $\x_f$ in time $\tau$, and the value of $\tau$ is then
determined (in terms of $\x_0$ and $\x_f$) by the constraint
equation, and similarly for $\y(t)$. The
decoherence functional is again essentially a sum over pairs of
classical paths for the system variables, the path
from $\x_0$ to $\x_f$, and the path from $\y_0$ to $\x_f$.

\subhead{\bf 6(B). Calculation of the Influence Functional}

We may now calculate the influence functional with the semiclassical
approximation for system variables inserted. The influence
functional may be written,
$$
F (\x_f, \x_0, \y_0) = \int d^n q_f \ \phi (\q_f, \x_f, \x_0)
\phi^* (\q_f, \x_f, \y_0)
\eqno(6.12)
$$
where
$$
\phi ( \q_f, \x_f, \x_0 )
= \int d^n q_0 \ g ( \q_f, \x_f | \q_0, \x_0 ) \ \chi (\q_0, \x_0 )
\eqno(6.13)
$$
Here, we have introduced the propagator for the environment variables
along the system classical trajectory from $\x_0$ to $\x_f $,
$$
g ( \q_f, \x_f | \q_0, \x_0 )
= \int \D \q(t)
\exp \left( i S_{\E}^{\tau} [\x(t), \q (t) ] \right)
\eqno(6.14)
$$
The environment state $\chi$ may be normalized according to
$$
\int d^n q \ \left| \chi (\q, \x) \right|^2 = 1
\eqno(6.15)
$$
for all $\x$. Since $g$ propagates $\chi$ unitarily along
a fixed system trajectory, it follows that
$$
\int d^n q_f \ \left| \phi (\q_f, \x_f, \x_0 \right|^2 =1
\eqno(6.16)
$$
This means that the influence functional satisfies
$$
\left| F( \x_f, \x_0, \y_0 ) \right|^2 \le 1
\eqno(6.17)
$$
with equality when $\x_0 = \y_0 $, indicating
that the influence functional is peaked about $\x_0 = \y_0$,
which is the decoherence effect we need.

The influence functional is difficult to evaluate in
general. It can be evaluated exactly when both $g$ and $\chi$ are
Gaussian. It is also effectively Gaussian in form if we have a
large number of oscillators in the environment, for then the
many-oscillator influence functional is essentially the
original one, (6.9), raised to a high power. This strongly
enhances the peaking about $\x_0 = \y_0$. A simple and reasonably
general form for the influence functional may therefore be obtained
by expanding about $\x_0 = \y_0$, and assuming either a Gaussian
form, or a large number of oscillators (or both).

We again use the coordinates $\X_0, \v$ defined in Eq.(4.11). We have
$$
\phi (\q_f, \x_f, \x_0) = \phi (\q_f, \x_f, \X_0)
+ \half v^a \ \d_a  \phi (\q_f,\x_f, \X_0)
+ {1 \over 8} v^a v^b \d_a \d_b \phi (\q_f, \x_f, \X_0)
+ \cdots
\eqno(6.18)
$$
where $\d_a = { \d / \d X_0^a} $.
Inserting in the influence functional, and also introducing
the notation $\phi_0 = \phi (\q_f, \x_f, \X_0)$, we get
$$
\eqalignno{
F = & \ 1 + \half v^a \int d^n q_f \left( \phi_0^* \d_a \phi_0
- \phi_0 \d_a \phi_0^* \right)
\cr
& + {1 \over 8} v^a v^b \int d^n q_f
\left( \phi_0^* \d_a \d_b \phi_0 + \phi_0 \d_a \d_b \phi_0^*
- 2 \d_a \phi_0 \d_b \phi_0^* \right)
\cr
& + \cdots
&(6.19) \cr }
$$
By differentiating the normalization of $\phi$, (6.16), we see that
$$
\int d^n q_f \ \phi_0 \d_a \phi_0^* = - \int d^n q_f \ \d_a \phi_0
\phi_0^*
\eqno(6.20)
$$
and
$$
\int d^n q_f \ \left( \phi_0 \d_a \d_b \phi_0^*
+ \phi_0^* \d_a \d_b \phi_0 \right)
= - \int d^n q_f \left( \d_a \phi_0 \d_b \phi_0^* +
\d_b \phi_0 \d_a \phi_0^* \right)
\eqno(6.21)
$$
Using these relations,
we may now write the influence functional as
$$
\eqalignno{
F & = \ 1 + i v^a \Gamma_a - \half v^a v^b \Sigma_{ab} + \cdots
\cr
& \approx \exp \left( i v^a \Gamma_a - \half v^a v^b \left(
\Sigma_{ab} - \Gamma_a \Gamma_b \right) \right) + \cdots
&(6.22) \cr }
$$
The coefficients $\Gamma_a $ and $\Sigma_{ab}$ are given by
$$
\eqalignno{
\Gamma_a (\x_f, \X_0 ) =&  \ {i  \over 2}
\int d^n q_f \ \left( \phi_0 \d_a \phi_0^* - \phi_0^* \d_a \phi_0 \right)
\cr
\Sigma_{ab} (\x_f, \X_0 ) =& \int d^n q_f \ \d_a \phi_0 \d_b \phi_0^*
&(6.23) \cr }
$$
The approximation of writing $F$ as a Gaussian becomes exact
when $g$ and $\chi$ are Gaussian, and is also true approximately
when there are a large number of oscillators in the environment.
We have therefore obtained the influence functional, as required.

\subhead{\bf 6(C). Reparametrization Invariance in the Influence
Functional}

The form of the influence functional indicates, as expected, that
there is a suppression of interference for paths with
$\x_0 \ne \y_0$. This is the usual decoherence effect. However,
the situation in the reparametrization-invariant theory
considered here is not so simple. We expect that the decoherence
functional depends, in some sense,
only on reparametrization-invariant quantities,
and in the semiclassical approximation used here, this means
it depends on entire classical paths (rather than individual
points). Differently put, we do not expect (or need) the
destruction of interference for points $\x_0$, $\y_0$ lying
on the {\it same} classical path (connecting each to $\x_f$),
since these points are effectively equivalent.
What we expect is that the influence functional will not be
exponentially small when $ \x_f, \x_0, \y_0$ lie along
a single classical path. We must therefore
see how reparametrization invariance is expressed in
the influence functional.

We first note that
$$
\phi_0
= \int d^n q_0 \ g ( \q_f, \x_f | \q_0, \X_0 ) \ \chi (\q_0, \X_0 )
\eqno(6.24)
$$
and let us see how this quantity varies with $\X_0$. If this
were the usual non-relativistic quantum mechanics, with propagation
from initial time $t_0$ to final time $t_f$, then $\phi_0$ would
in fact be independent of $t_0$. We expect a similar property
here, that is, that $\phi_0$ is constant (as a function of $\X_0$)
along a certain vector field. The initial state
$\chi (\q_0, \X_0)$ obeys the Schr\"odinger equation
$$
i \nabla_0 S (\X_0) \cdot \nabla_0 \chi = H_{_\E} (\q_0, \x_f, \X_0 ) \chi
\eqno(6.25)
$$
The propagator $g$ on the other hand, obeys Schr\"odinger
equations with respect to both the final and initial points,
$$
\eqalignno{
i \nabla_f A (\x_f, \X_0 ) \cdot \nabla_f g & = H_{_\E}(\q_f, \x_f, \X_0 )  g
&(6.26) \cr
i \nabla_0 A (\x_f, \X_0 ) \cdot \nabla_0 g & = H_{_\E} (\q_0, \x_f, \X_0 )  g
&(6.27) \cr}
$$
(Note that the expected minus sign in the Schr\"odinger equation
with respect to the initial point is already contained through
the fact that $ \nabla_0 A $ is {\it minus} the initial momentum, as in
Eq.(4.6)). Now the point is here that $g$ and $\chi$ obey
different Schr\"odinger equations, so at this stage, $\phi_0$ does
not obviously have any constant directions in $\X_0$ -- neither
$\nabla_0 S \cdot \nabla_0 \phi_0$ nor $\nabla_0 A \cdot \nabla_0
\phi_0$ are zero.

However, as we saw in the Section 4 (without environment),
the path integral enforces the condition $ \p = - \nabla_0 A $. We
anticipate that
this condition is approximately enforced with the environment in
place. Furthermore, the initial Wigner function for a WKB state is
of the approximate form,
$$
W (\p, \X_0 ) = \left| C(\X_0) \right|^2 \ \delta ( \p - \nabla_0 S (\X_0) )
\eqno(6.28)
$$
It follows that the sum over paths is dominated
by configurations for which
$$
\nabla_0 A (\x_f, \X_0 ) \approx - \nabla_0 S
\eqno(6.29)
$$
This means that the trajectories of $S$ are the same as the
classical trajectories from $\X_0$ to $\x_f$.
From this it is then easy to show that
$$
\nabla_0 S (\X_0) \cdot \nabla \phi_0 \approx 0
\eqno(6.30)
$$
(essentially for the same reason that the analogous non-relativistic
version is independent of the initial time $t_0$).
In the influence functional, two neighbouring points
$\x_0$, $\y_0$ on the same classical trajectory
have $ \v = \x_0 - \y_0 $ proportional to $\nabla_0 S$.
It follows that
$$
v^a \Gamma_a = 0, \quad v^a \Sigma_{ab} = 0
\eqno(6.31)
$$
which means that the influence functional does not suppress
interference between points on the same trajectories, only between
points on different trajectories. That is, when the condition $\p = -
\nabla_0 A$ is true, we get the expected result that the influence
functional is a function only of entire trajectories, and not of the
individual points along those trajectories.

\subhead{\bf 6(D). Decoherence and the Evaluation of the v integral}

The off-diagonal terms of the decoherence functional may now be written
$$
\eqalignno{
D (\Delta , \bar \Delta)  &= \int d^n x_f d^n X_0 d^n v d^n p
\ \theta \left(
\int_{-\infty}^{\infty} dt \ f_{\Delta} (\x_0^f (t) ) -\e \right)
\cr
& \times
\ \left( 1 - \theta \left( \int_{-\infty}^{\infty} dt \ f_{\Delta} (\y_0^f (t) )
-\e \right) \right)
\cr
& \times P (\x_f, \X_0 + \half \v) P^* (\x_f, \X_0 - \half \v)
\cr & \times
\ \exp \left( i \left( \nabla_0 A (\x_f,
\X_0) + \p \right) \cdot \v + O (\v^3) \right) \ W (\p, \X_0)
\cr & \times
\exp \left( i v^a \Gamma_a - \half v^a v^b \sigma_{ab}
\right) &(6.32) \cr }
$$
where
$$
\sigma_{ab} = \Sigma_{ab} - \Gamma_a \Gamma_b
\eqno(6.33)
$$
and we have again introduced the variables $\X_0$, $\v$
in the exponential part.
The decoherence functional is a sum over pairs of classical
paths, one set of paths intersecting $\x_0, \x_f$ and
passing through $\Delta$ at any stage along the path, the
other set of paths intersecting $\y_0, \x_f$ and
never passing through $\Delta$ at any stage along the
path (see Fig.1). It is easily seen that for this particular
coarse graining, in which we are interested in paths that
either pass or do not pass through the region $\Delta$,
we do not in fact encounter the situation discussed in Section
6(C), in which $\x_0, \y_0, \x_f$ lie along the same
classical trajectory. That is, the coarse graining is such
that $\v$ is never proportional to $\nabla_0 A$, and
the potentially singular situation (6.31) does not arise.
The influence functional therefore does its job of
suppressing the contribution from non-zero values
of $\v$ to a decoherence width determined by
the inverse of the non-zero eigenvalues of
$\sigma_{ab}$.
If the size of the coarse graining region $\Delta $
is greater than this width then the off-diagonal terms
of the decoherence functional are approximately
zero.

We therefore have approximate decoherence and we may
examine the probability for passing through $\Delta$,
which is
$$
\eqalignno{
p (\Delta )  &= \int d^n x_f d^n X_0 d^n v d^n p
\ \theta \left(
\int_{-\infty}^{\infty} dt \ f_{\Delta} (\X_0^f (t) ) -\e \right)
\cr
& \times P (\x_f, \X_0 + \half \v) P^* (\x_f, \X_0 - \half \v)
\cr & \times
\ \exp \left( i \left( \nabla_0 A (\x_f,
\X_0) + \p \right) \cdot \v + O (\v^3) \right) \ W (\p, \X_0)
\cr & \times
\exp \left( i v^a \Gamma_a - \half v^a v^b \sigma_{ab}
\right) &(6.34) \cr }
$$
where $\v$ has been set to zero in the $\theta$-function.
We are now summing over pairs of classical paths which
{\it both} pass through the region $\Delta$, so now we
do have the possibility of $\x_0, \y_0, \x_f$ lying along
the same path, and hence the matrix $\sigma_{ab}$ is potentially
singular, by (6.31). This means that some care is
necessary in the $\v$ integral.

If we formally carry out the integral over $\v$, we get,
$$
\eqalignno{
p_{\Delta}
=&
\int d^n x_f d^n X_0 d^n p \ \ \theta \left(
\int_{-\infty}^{\infty} dt \ f_{\Delta} (\X_0^f (t) ) -\e \right)
\cr
& \times \left| P (\x_f, \X_0 ) \right|^2 W ( \p, \X_0 )
\cr & \times
\exp \left( - \half ( \nabla_0 A (\x_f, \X_0 ) + \p + \Gamma)^a
\sigma_{ab}^{-1} ( \nabla_0 A (\x_f, \X_0 ) + \p + \Gamma)^b
\right)
&(6.35) \cr }
$$
Changing integration variables from $\x_f $ to
$\p_0 =- \nabla_0 A (\x_f, \X_0)$
This is conveniently written
$$
p_{\Delta} = \int d^n p_0 d^n X_0
\ \theta \left(
\int_{-\infty}^{\infty} dt \ f_{\Delta} (\X^{cl} (t) ) -\e \right)
\tilde W (\p_0, \X_0)
\eqno(6.36)
$$
where we have defined the smeared Wigner function
$$
\tilde W (\p_0 , \X_0)
= \int d^n p
\ \exp \left( - \half ( \p_0 - \p - \Gamma)^a
\sigma_{ab}^{-1} ( \p_0 - \p - \Gamma)^b
\right)
\ W (\p, \X_0)
\eqno(6.37)
$$
Again the prefactors $P$ drop out in the change of
integration variables, as in Section 4. This smearing of the
Wigner function represents environmentally induced fluctuations
about the classical evolution, and the small additional term
$\Gamma$ in the exponent represents the back reaction
of the environment on the classical equations of motion.

As stated, the results (6.31) suggest that the matrix $\sigma_{ab}$
is singular, but it is easy to see the significance of this.
When the matrix is non-singular, the $\v$ integral produces
a Gaussian peak about $\p_0 = \p + \Gamma$, which, as we have
seen represents fluctuations about classical evolution.
If the matrix is singular in a certain direction, it is easy to see
from (6.34) that the $\v$ integral in this direction will produce
a $\delta$-function, instead of a Gaussian. It will still be
peaked about the same configuration, but there are no fluctuations
in that direction.

The result for the probabilities therefore approximately
coincides with the classical result Eq.(2.10). We have concentrated
on the case in which $\Delta $ is a single region of
configuration space, but the result straightforwardly
generalizes to the case in which $\Delta $ consists of
a series of regions $\Delta_1, \Delta_2, \cdots$. The
above result then shows that the probability is peaked
when the series of regions lies along a classical
path (plus the environmental effects of a small back reaction
and small fluctuations).

%[MORE ON FLUCTUATIONS. BACK REACTION EQUATIONS].

\head{\bf 7. Superposition States}

The calculations of Sections 4 and 6 concerned only single
WKB states of the form (6.4). It is therefore important to reconsider
the decoherence calculation of Section 6 for the more general case
of a superposition of WKB states,
$$
\Psi = \Psi_1 + \Psi_2 = C_1 e^{i S_1} \chi_1 + C_2 e^{ i S_2} \chi_2
\eqno(7.1)
$$
This turns out in fact to be quite straightforward, mainly because
similar calculations (involving the reduced density matrix, not
the decoherence functional) have already been done.

Inserting Eq.(7.1) in the decoherence functional, we obtain,
a result of the form
$$
D = D_{11} + D_{12} + D_{21} + D_{22}
\eqno(7.2)
$$
where, in an obvious notation,
$D_{11}$ is the decoherence functional with initial density
matrix $ | \Psi_1 \ra \la \Psi_1 | $, $D_{12}$ is the decoherence
functional with the operator $ | \Psi_1 \ra \la \Psi_2 | $
in the initial state slot, and so on. Clearly the analysis of
$D_{11}$ and $D_{22}$ is identical to the case considered
already -- we get decoherence, and probabilities given in terms
of the Wigner functions of $\Psi_1$ and $\Psi_2$. Hence these
two terms correspond to a statistical mixture of the two initial
states.

The interesting terms are $D_{12}$ and $D_{21}$ ($=D_{12}^*$),
which correspond to interferences between different WKB branches.
From Section 6, we see that
$$
\eqalignno{
D_{12} (\a, \a') = &
\int_{-\infty}^{\infty} d \tau
\int_{-\infty}^{\infty} d \tau'
\ e^{ - i E \tau } \ e^{  i E' \tau' }
\cr & \times
\int_{\a} \D \x(t) \ \int_{\a'} \D \y(t)
\exp \left( i S_0^\tau [\x(t)] - i S_0^{\tau'}
[\y(t)] \right)
\cr & \times
\ F_{12} [ \x(t), \y(t), \tau, \tau' ]
\ C_1 (\x_0) e^{i S_1(\x_0)} \ C_2^* (\y_0) e^{ - i S_2 (\y_0) }
&(7.3) \cr}
$$
where
$$
\eqalignno{
F_{12} [ \x(t), \y(t), \tau, \tau' ] = & \int \D \q(t) \D \r (t)
\exp \left( i S_{\E}^{\tau} [\x, \q] - i S_{\E}^{\tau'} [\y, \r]
\right)
\cr &
\quad \quad \times \ \chi_1 (\x_0, \q_0 ) \chi_2^* (\y_0, \r_0)
&(7.4) \cr}
$$
(As in Section 6(A) we should then replace the class operators
with their modified version and go to the semiclassical limit).
Now it is easy to see that the influence functional is the overlap
of the two initial states, but with each unitarily evolved along
two different trajectories. That is, in the semiclassical
approximation for the system,
$$
F_{12} = \la \chi_{2} (\y_0) | U^{\dag} (\x_f, \y_0 ) U( \x_f,
\x_0 ) | \chi_1 (\x_0) \ra
\eqno(7.5)
$$
where $U(\x_f, \x_0)$ denotes the unitary evolution of the environment
states along the system classical trajectory from $\x_0$ to $\x_f$.

Clearly $ |F_{12} |^2 \le 1 $, and because $F_{12}$ is an overlap
between a pair of states it will typically be such
that $F_{12} $ is strictly less than $1$. In this case, when
raised to a high power, as happens when we take a large number
of oscillators in the environment, we will get a very strong
suppression of terms with $\x_0 \ne \y_0$. In particular,
even when $\x_0 = \y_0$, we get
$$
F_{12} = \la \chi_2 (\x_0) | \chi_1 ( \x_0) \ra
\eqno(7.6)
$$
which will be less than $1$, quite simply because
$\chi_1$ and $\chi_2$
are different states. We therefore find that $D_{12}$ and
$D_{21}$ are much smaller than the diagonal terms
$D_{11}$, $D_{22}$.
This destruction of interference between
WKB states therefore comes about for essentially the same reason
that the corresponding off-diagonal terms of the density matrix
are very small, as discussed previously [\cite{Hal8,Kie2,Pad,Kie3}].

It should be noted, that in Eq.(7.5), it could
in fact happen that $F_{12} = 1 $, as a result of a careful choice of
$\chi_1 \ne \chi_2$ together with a suitable choice of
$\x_0 \ne \y_0$, in particular, if
$$
U( \x_f, \x_0 ) | \chi_1 (\x_0) \ra
= U (\x_f, \y_0 ) | \chi_2 (\y_0) \ra
\eqno(7.7)
$$
The point here, however, is that this becomes very unlikely
with a large environment. With a large collection of oscillators
in the environment, the environment states are a
tensor product over $A$ of states $| \chi_1^A (\x_0) \ra$,
for example, and then
$$
F_{12} = \prod_A \la \chi^A_{2} (\y_0) | U^{\dag} (\x_f, \y_0 ) U( \x_f,
\x_0 ) | \chi^A_1 (\x_0) \ra
\eqno(7.8)
$$
As the size of the environment goes to infinity, the possibility
of Eq.(7.8) being exactly $1$ becomes negligible.

It is also of interest to look at the special case in which the
wave function Eq.(7.1) is real (as is the case in the no-boundary
wave function of Hartle and Hawking), so that $\Psi_2 = \Psi_1^*$,
and $ \chi_2 = \chi_1^* $. When $\x_0 = \y_0$, we then have that
$$
\left| F_{12} \right|^2 =  \left| \int d \q_0 \ \chi_1^2 (\x_0, \q_0)
 \right|^2
\eqno(7.9)
$$
Since $\chi_1$ is generally complex (it obeys the complex Schr\"odinger
equation (6.7)), the right-hand side of Eq.(7.9)
will again be less than $1$, so the argument still goes through
[\cite{Kie3}]. The argument fails if $\chi_1$ is real. But then
it would have to be an eigenstate of the environment Hamiltonian
for all values of $\x_0$, and this would not lead to decoherence,
so we may disregard this case.

% [CHECK KIEFER REFERENCE].

\head{\bf 8. Summary and Discussion}

We have studied the quantization and interpretation of simple
timeless models described by an equation of the type (1.1). In
particular, we studied the question, what is the probability that
the system passes through a region $\Delta$ of configuration space
without reference to time?

We obtained the classical answer to this problem, in three
different forms, in terms of a classical phase space distribution
function $w(\p, \x)$, satisfying $\{ H, w  \} = 0$, the analogue
of (1.1). This function needs some care in normalization since it
is constant along the (possibly infinite) classical orbits. A very
useful step in the classical case was the introduction of the
phase space quantity Eq.(2.7), which is $\delta$-function peaked
on the classical path and also has vanishing Poisson bracket with
the Hamiltonian, so is an observable. This quantity assists in
understanding some aspects of the quantum theory.

We constructed the decoherence functional, following the general
scheme of Refs.[\cite{Har3,HaMa}], using the induced inner
product. Although the general scheme has been presented
previously, a key part of our contribution to this area is the
explicit identification (in the semiclassical approximation) of
the class operator Eq.(4.8) {\it satisfying the constraint
everywhere} describing histories restricted to pass through a
region $\Delta$ of configuration space. Having made this
identification, a major part of our work was to show that the
decoherent histories approach then reduces, approximately, to the
corresponding classical result, but with the classical phase space
distribution function $w(\p, \x)$ replaced by the Wigner function
$W(\p, \x)$ of the quantum theory. We also explored the
decoherence and probabilities for a system of harmonic oscillators
using the timeless coherent states, in terms of which the analysis
is particularly transparent and fully agrees with intuitive
expectations.

In brief, therefore, we have shown that heuristic classically
inspired notions of interpretation for simple timeless models may
in fact be derived from the decoherent histories analysis of such
models. This result is by no means unexpected, but the key aspects
of the derivation are the elucidation of the role of the
constraint and the related reparametrization invariance in the
construction of both the classical and quantum results. Furthermore,
the complete absence of a time parameter is not an obstruction
to quantization.

There are a number of issues which the present work has generated
and will be discussed in a later publication, but we mention them
briefly below.

First of all, the main difficulty in computing the decoherence
functional for our chosen coarse graining is the calculation of
the modified class operators. Even before modification class
operators of the type Eq.(3.21) are difficult to calculate
(typically they can only be obtained exactly in the very simple
situations where the method of images may be used). The suggested
scheme for constructing modified operators obeying the constraint
has not yet been explored fully. Here, we have constructed
physically plausible modified class operators in the semiclassical
approximation, obtaining full agreement with the classical
results. Some exact modified class operators for simple coarse
grainings of the relativistic particle have been constructed in
Ref.[\cite{HaTh}], but it is not yet clear how general those
results are. Hence a more detailed investigation of these modified
class operators is called for. (We note in passing that, from the
simple examples in which the modified class operators have been
calculated, their calculation does in fact appear to be
considerably easier than the original ones, Eq.(3.21)).

Second, we have assumed (except in Section 5) that both the
initial states and the propagators are in the oscillatory regime.
This means in the propagator that we assume the dominant
contribution comes from real configurations (rather than Euclidean
or complex ones). Many interesting models in quantum cosmology
have a Euclidean region, corresponding, for example, to
``tunneling from nothing''. It is not immediately clear how the
semiclassical calculation of Section 4 is modified to include this
case, the main difficulty being understanding what the class
operators are. This case is therefore probably related to the
question of a more general formula for the modified class
operators.

Third, it is generally understood that decoherence of histories is
related to the existence of ``records''[\cite{GeHa,Hal6}]. This
means that it is possible to find a projection operator $R_{\au}$
which is perfectly correlated with the class operators $C_{\au}$
in terms of which the probabilities may be written,
$$
p(\au ) = {\rm Tr} \left( C_{\au} \rho C_{\au} \right) = {\rm Tr}
\left( R_{\au} \rho \right)
\eqno(8.1)
$$
In the case of a non-relativistic model where decoherence is
produced by an environment, it is possible to explicitly identify
the environmental variables which store the information about the
system [\cite{Hal6}]. It would be very desirable to do this in the
timeless case considered here. It seems likely that the variables
are very similar to the case of Ref.[\cite{Hal6}], but the
interesting question is the role of reparametrization invariance
in this situation, and whether the records are closely related to
observables in the operator approach.

A fourth issue concerns the connection between the decoherent
histories analysis considered here and the master equation for the
reduced density operator (in the case where decoherence is
produced by an environment). It would be of interest to see if the
discussion of decoherence and probabilities can be re-expressed in
the simpler language of the density operator, as it sometimes can
in non-relativistic decoherence models. This is currently under
investigation [\cite{Tho}].

These and related issues will be taken up in future publications.

\head{\bf Acknowledgements}

We are grateful to Jim Hartle, John Klauder and
Don Marolf for useful conversations.
JT was funded by Evangelisches Studienwerk.

\endpage

\epsfbox{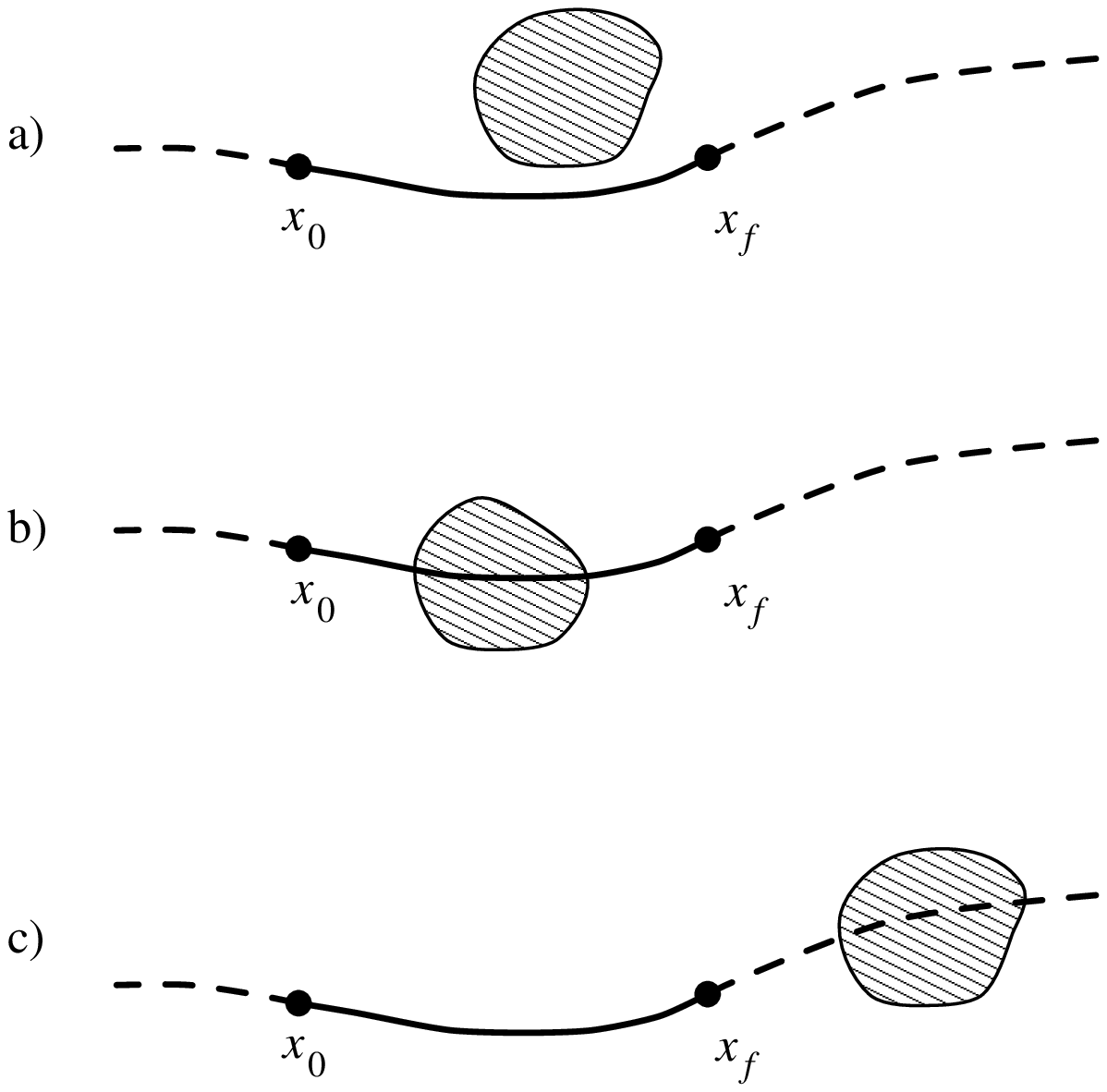}

\subhead{\bf Figure 1}

The rewritten classical probability Eq.(2.13) in terms of a sum
over initial and final points $\x_0$ and $\x_f$. The probability
for not entering $\Delta$ is a sum over paths as in case (a).
The probability for entering $\Delta$ includes a sum
over classical paths in which $\Delta$ lies between the initial
and final points, as in case (b).
But, to agree with the phase space form of the
result Eq.(2.10), it must also include a sum over initial and
final points for which $\Delta $ does not lie between them,
as in case (c). This figure also applies to the semiclassical
propagator Eq.(4.8) in the quantum case.

\endpage

\references

\def\pr{{\sl Phys. Rev.\ }}

\def\prep{{\sl Phys. Rep.\ }}
\def\jmp{{\sl J. Math. Phys.\ }}

\def\np{{\sl Nucl. Phys.\ }}

\def\annp{{\sl Ann. Phys. (N.Y.)\ }}
\def\cqg{{\sl Class. Quant. Grav.\ }}

\refis{Ash} For a nice review see, C.Rovelli, gr-qc/9710008,
{\it Loop quantum gravity}.

%\refis{asy} See, for example, the collection of articles in {\it
%Physical Origins of Time Asymmetry}, edited by J.J.Halliwell,
%J.Perez-Mercader and W.Zurek (Cambridge University Press,
%Cambridge, 1994). See also, H.D.Zeh, {\it The Physical Basis of the
%Direction of Time}, third edition (Springer-Verlag, 1999)
%(and the associated webpage www.time-direction.de), and also
%K.Ridderbos, {\sl Studies in History and Philosophy of Modern Physics}
%{\bf 30}, 41(1999).

\refis{AuKi} A.Auerbach and S.Kivelson, \np {\bf B257}, 799 (1985).
%The path decomposition expansion and multidimensional tunneling
See also, J.J.Halliwell, {\sl Phys.Lett.} {\bf A207}, 237--242 (1995).
%``An Operator Derivation of the Path Decomposition
%Expansion''.

\refis{BaJ} N.Balazs and B.K.Jennings, \prep {\bf 104}, 347 (1984),
M.Hillery, R.F.O'Connell, M.O.Scully and E.P.Wigner, \prep {\bf
106}, 121 (1984).

\refis{Bar1} J.Barbour, \cqg {\bf 11}, 2853 (1994).
% The timelessness of quantum gravity. I. The evidence
% the classical theory.

\refis{Bar2} J.Barbour, \cqg {\bf 11}, 2875 (1994).
% The timelessness of quantum gravity. II. The appearance of
% dynamics in static configurations.

\refis{Bar3} J.Barbour, {\it The End of Time: The Next Revolution
in our Understanding of the Universe} (Weidenfeld and Nicholson, 1999).

\refis{BuIs} J.Butterfield and C.J.Isham, gr-qc/9901024,
{\it On the emergence of time in quantum gravity}.

\refis{Bar} A.O.Barvinsky, {\sl Phys.Rep.} {\bf 230}, 237 (1993);
{\sl Phys.Lett} {\bf B241}, 201 (1990).

\refis{CrHa} D.Craig and J.B.Hartle, unpublished.
preprint UCSBTH-94-47 (1998).
% Generalized quantum theory of Bianchi IX Cosmologies.

%\refis{CoHe} K.Hepp, {\sl Helv.Phys.Acta} {\bf 45}, 237 (1972).
% More recent developments?

\refis{DeW} B.DeWitt, in {\it Gravitation: An Introduction to
Current Research}, edited by L.Witten (John Wiley and Sons, New
York, 1962).
% The quantization of geometry.

\refis{Gar} C.W.Gardiner, {\it Quantum Noise} (Springer-Verlag,
Berlin, 1991).

\refis{GeHa} M.Gell-Mann and J.B.Hartle, in {\it Complexity,
Entropy and the Physics of Information, SFI Studies in the
Sciences of Complexity}, Vol. VIII, W. Zurek (ed.) (Addison
Wesley, Reading, 1990); and in {\it Proceedings of the Third
International Symposium on the Foundations of Quantum Mechanics in
the Light of New Technology}, S. Kobayashi, H. Ezawa, Y. Murayama
and S. Nomura (eds.) (Physical Society of Japan, Tokyo, 1990);
%{\it Quantum Mechanics in the Light of Quantum Cosmology.}
{\sl Phys.Rev.} {\bf D47}, 3345 (1993).

\refis{Gri} R.B.Griffiths, {\sl
J.Stat.Phys.} {\bf 36}, 219 (1984); {\sl Phys.Rev.Lett.} {\bf 70},
2201 (1993); {\sl Am.J.Phys.} {\bf 55}, 11 (1987);
J.B.Hartle, in {\it Quantum Cosmology and Baby
Universes}, S. Coleman, J. Hartle, T. Piran and S. Weinberg (eds.)
(World Scientific, Singapore, 1991);
J.J.Halliwell, in {\it
Fundamental Problems in Quantum Theory}, edited by D.Greenberger
and A.Zeilinger, Annals of the New York Academy of Sciences, Vol
775, 726 (1994). For further developments in the decoherent
histories approach, particularly adapted to the problem of
spacetime coarse grainings, see C. Isham, \jmp {\bf 23}, 2157
(1994);
%{\it Quantum Logic and the Histories Approach to Quantum Theory.}
C. Isham and N. Linden, \jmp {\bf 35}, 5452 (1994); {\bf 36}, 5392
(1995).

\refis{Hal1} J.J.Halliwell, in, {\it Proceedings of the 13th
International Conference on General Relativity and Gravitation},
edited by R.J.Gleiser, C.N.Kozameh, O.M.Moreschi
(IOP Publishers, Bristol,1992). (Also available as
the e-print gr-qc/9208001).
% The Interpretation of Quantum Cosmological Models

\refis{Hal2} J.J.Halliwell, {\sl Phys.Rev.} {\bf D64}, 044008 (2001).
%``Trajectories for the Wave Function of the Universe from a
%Simple Detector Model'', % gr-qc/0008046.

\refis{Hal3} J.J.Halliwell, \pr {\bf D38}, 2468 (1988).
% Derivation of the Wheeler-DeWitt equation
% from a Path Integral for Minisuperspace Models.

%\refis{Hal4} This detector model was used in a simple
%non-relativistic context by J.J.Halliwell, \pr {\bf D60}, 105031
%(1999).
% ``Somewhere in the Universe: Where is the Information Stored when
% Histories Decohere?'', quant-ph/9902008, Imperial preprint
% TP/98-99/29 (1999).
%Some subsequent developments of the Coleman-Hepp model are
%H.Nakazato and S.Pascazio, \prl {\bf 70}, 1 (1993); \pr {\bf A48},
%1066 (1993); R.Blasi, S.Pascazio, S.Takagi, \pr {\bf A250}, 230 (1998).

%\refis{Hal5} J.J.Halliwell, {\sl Prog.Theor.Phys.} {\bf 102}, 707 (1999).

\refis{Hal6} J.J.Halliwell, {\sl Phys.Rev.} {\bf D60}, 105031 (1999).
%``Somewhere in the Universe: Where is the Information
%Stored when Histories Decohere?'',
%Imperial/TP/98--99/29. quant-ph/9902008.

\refis{Hal7} J.J.Halliwell, {\sl Phys.Rev.} {\bf D36}, 3626 (1987).
%``Correlations in the Wave Function of the Universe",

\refis{Hal8} J.J.Halliwell, , {\sl Phys.Rev.} {\bf D39}, 2912 (1989).
%``Decoherence in Quantum Cosmology"

\refis{HalH} J.J.Halliwell and J.B.Hartle, {\sl Phys.Rev.}
{\bf D43}, 1170 (1991).
% ``Wave Functions Constructed from an Invariant
% Sum-Over-Histories Satisfy Constraints",

\refis{HaOr} J.J.Halliwell and M.E.Ortiz, {\sl Phys.Rev.} {\bf D48}, 748 (1993).

\refis{HaTh} J.J.Halliwell and J.Thorwart, {\sl Phys.Rev.} {\bf D64}, 124018 (2001).
%``Decoherent histories analysis of the relativistic particle''
%(with J.Thorwart),

%\refis{HaKu} J.B.Hartle and K.Kuchar, \pr {\bf D34}, 2323 (1986).
% Path integrals in parameterized theories: the free relativistic
% particle

\refis{Har1} J.B.Hartle, \pr {\bf D37}, 2818 (1988).
% Quantum kinematics of spacetime. I. Non-relativistic theory.

\refis{Har2} J.B.Hartle, \pr {\bf D38}, 2985 (1988).
% Quantum kinematics of spacetime. II. A model quantum cosmology
% with real clocks.

\refis{Har3} J.B.Hartle, in {\it Proceedings of the 1992 Les Houches
School, Gravity and its Quantizations}, edited by B.Julia and
J.Zinn-Justin (Elsevier Science B.V. 1995), also
available as the e-print, gr-qc/9304006.
% Spacetime quantum mechanics and the quantum mechanics of
% spacetime

\refis{HaMa} J.B.Hartle and D.Marolf,  \pr {\bf D56}, 6247 (1997).
% Comparing Formulations of Generalized Quantum Mechanics for
% Reparametrization-Invariant Systems

\refis{HaHa} J.B.Hartle and S.W.Hawking, \pr {\bf D28}, 2960
(1983).

%\refis{Haw} S.W.Hawking, \pr {\bf D32}, 2489 (1985).
% The arrow of time in cosmology.

\refis{Haw} S.W.Hawking, {\sl Nucl.Phys.} {\bf B239}, 257 (1984).
% The quantum state of the universe.

\refis{HaPa} S.W.Hawking and D.N.Page, {\sl Nucl.Phys.}
{\bf B264}, 185 (1986);
%Operator ordering and the flatness of the universe.
{\bf B298}, 789 (1988).
%How probable is inflation?

\refis{Ish} C.J.Isham, gr-qc/9210011,
{\it Canonical quantum gravity and the problem of time}.

%\refis{KaNa} Y.Kazama and R.Nakayama, \pr {\bf 32}, 2500 (1985).
% Wave packet in quantum cosmology.

%\refis{Kie} C.Kiefer, \pr {\bf D38}, 1761 (1988).
% Wave packets in minisuperspace.

\refis{Kie2} C.Kiefer, {\sl Class.Quant.Grav.} {\bf 4}, 1369 (1987).
% CONTINUOUS MEASUREMENT OF MINISUPERSPACE VARIABLES BY HIGHER
% MULTIPOLES.

\refis{Kie3} C.Kiefer, {\sl Phys.Rev.} {\bf D46}, 1658 (1992).

\refis{Kla} J.Klauder, Ann. Phys. (NY) {\bf 254}, 419 (1997),
(quant-ph/9604033) ; Nucl.Phys. {\bf B547},397, 1999,
(hep-th/9901010); hep-th/0003297.

\refis{Kuc} K.Kuchar, in {\it Conceptual Problems of Quantum
Gravity}, edited by A.Ashtekar and J.Stachel (Boston, Birkhauser,
1991); and in {\it Proceedings of the 4th Canadian Conference on
General Relativity and Relativistic Astrophysics}, edited by
G.Kunstatter, D.E.Vincent and J.G.Williams (World Scientific, New
Jersey, 1992). See also the e-print gr-qc/9304012,
{\it Canonical quantum gravity}.

\refis{Mar1} D.Marolf, \cqg {\bf 12}, 1199 (1995).
% Quantum observables and recollapsing dynamics

\refis{Mar2} D.Marolf, \pr {\bf D53}, 6979(1996);
% Path Integrals and Instantons in Quantum Gravity
\cqg {\bf 12}, 2469 (1995);
% Almost Ideal Clocks in Quantum Cosmology: A Brief Derivation of Time
\cqg{\bf 12}, 1441 (1995).
% Observables and a Hilbert Space for Bianchi IX

\refis{Mot} N.F.Mott, {\sl Proc.Roy.Soc} {\bf A124}, 375 (1929).
% The wave mechanics of alpha-ray tracks.
See also J.S.Bell, {\it Speakable and Unspeakable in Quantum Mechanics}
(Cambridge University Press, Cambridge, 1987), and
A.A.Broyles, {\sl Phys.Rev.} {\bf A48}, 1055 (1993),
for further discussions of the Mott calculation.

\refis{Omn} R.Omn\`es, {\sl J.Stat.Phys.}
{\bf 53}, 893 (1988); {\bf 53}, 933 (1988); {\bf
53}, 957 (1988); {\bf 57}, 357 (1989); {\bf 62}, 841 (1991); {\sl
Ann.Phys.} {\bf 201}, 354 (1990); {\sl Rev.Mod.Phys.} {\bf 64},
339 (1992).

\refis {Pad} T.Padmanabhan, {\sl Phys.Rev.} {\bf D39}, 2924, (1989).
%Decoherence in the density matrix describing quantum three-geometries and the
%emergence of classical spacetime.

\refis{Rie} A.Ashtekar, J.Lewandowski, D.Marolf, J.Mourao and
T.Thiemann, {\sl J.Math.Phys.} {\bf 36}, 6456 (1995);
% Quantization of diffeomorphism invariant theories of connections with
% local degrees of freedom.
A.Higuchi, \cqg {\bf 8}, 1983 (1991).
% Quantum linearization instabilities of de Sitter spacetime 2.
D.Giulini and D.Marolf, \cqg {\bf 16}, 2489 (1999);
% A Uniqueness Theorem for Constraint Quantization
\cqg {\bf 16}, 2479 (1999).
% On the Generality of Refined Algebraic Quantization
F.Embacher, {\sl Hadronic J.} {\bf 21}, 337 (1998);
% Handwaving refined algebraic quantization.
N.Landsmann, {\sl J.Geom.Phys.} {\bf 15}, 285 (1995).
% Rieffel induction as generalized quantum Marsden-Weinstein
% reduction

\refis{Rov1} C.Rovelli, \pr {\bf 42}, 2638 (1990).
% Quantum mechanics without time: A model.

\refis{Rov2} C.Rovelli, \pr {\bf 43}, 442 (1991).
% Time in quantum gravity: An hypothesis.

\refis{Rov3} C.Rovelli, \cqg {\bf 8}, 297 (1991);
% What is observable in classical and quantum gravity
{\bf 8}, 317 (1991).
% Quantum references systems.

\refis{Sch} L.S.Schulman, {\it Techniques
and Applications of Path Integration} (Wiley-Interscience, New York, 1981),
Chapter 18.

\refis{Tat} V.I.Tatarskii, {\sl Sov.Phys.Usp} {\bf 26}, 311 (1983).

\refis{Tei} C.Teitelboim,
\pr {\bf D25}, 3159 (1983); {\bf 28}, 297 (1983);
{\bf 28}, 310 (1983).

\refis{Tho} J.Thorwart, in preparation.

\refis{Time} The literature on arrival and tunneling
times, and on time operators is vast. A selection is,
Y.Aharanov and D.Bohm, \pr {\bf 122}, 1649 (1961);
% Time in quantum theory and the uncertainty relation for time
Y.Aharanov, J.Oppenheim, S.Popescu, B.Reznik and
W.Unruh, quant-ph/9709031 (1997);
% Measurement of Time-of-Arrival in Quantum Mechanics.
G.R.Allcock, \annp {\bf 53}, 253 (1969);
{\bf 53}, 286 (1969); {\bf 53}, 311 (1969);
% The Time of Arrival in Quantum Mechanics
Ph.Blanchard and A.Jadczyk,
{\sl Helv.Phys.Acta.} {\bf 69}, 613 (1996);
% Time of events in quantum theory.
I.Bloch and D.A.Burba, \pr {\bf 10}, 3206 (1974);
% Presence of a particle in a given sacpe-time region and the
% continuous action of a particle dectector.
V.Delgado, preprint quant-ph/9709037 (1997);
R.Giannitrapani, preprint quant-ph/9611015 (1998);
N.Grot, C.Rovelli and R.S.Tate, \pr {\bf A54}, 46 (1996);
E.Gurjoy and D.Coon, {\sl Superlattices and
Microsctructures} {\bf 5}, 305 (1989);
A.S.Holevo, {\it Probabilistic and Statistical Aspects
of Quantum Theory} (North Holland, Amsterdam, 1982), pages 130--197;
A.Jadcyk, {\sl Prog.Theor.Phys.} {\bf 93}, 631 (1995);
% Particle Tracks, Events and Quantum Theory.
D.H.Kobe and V.C.Aguilera--Navarro,  \pr {\bf A50}, 933 (1994);
N.Kumar, {\sl Pramana J.Phys.} {\bf 25}, 363 (1985);
J.Le\'on, preprint quant-ph/9608013 (1996);
D.Marolf, \pr {\bf A50}, 939 (1994);
L.Mandelstamm and I.Tamm, {\sl J.Phys.} {\bf 9}, 249 (1945);
J.G.Muga, S.Brouard and D.Mac\'ias, \annp {\bf 240}, 351 (1995);
% Time of arrival in quantum mechanics.
J.G.Muga, J.P.Palao and C.R.Leavens, preprint quant-ph/9803087
(1987);
J.G.Muga, R.Sala and J.P.Palao, preprint quant-ph/9801043,
{\sl Superlattices and Microstructures} {\bf 23} 833 (1998);
C.Piron, in {\it
Interpretation and Foundations of Quantum Theory}, edited by
H.Newmann (Bibliographisches Institute, Mannheim, 1979);
M.Toller, preprint quant-ph/9805030 (1998);
H.Salecker and E.P.Wigner, \pr {\bf 109}, 571 (1958);
F.T.Smith, \pr {\bf 118}, 349 (1960);
E.P.Wigner, \pr {\bf 98}, 145 (1955).
See also the forthcoming volume,
{\it Time in Quantum Mechanics},
edited by J.G.Muga, R.Sala Mayato and I.L.Egususquiza
(to appear in 2002).

\refis{Whe} J.Whelan, \pr {\bf D50}, 6344 (1994);
% Spacetime alternatives in relativistic particle motion

\refis{YaT} N.Yamada and S.Takagi, {\sl Prog.Theor.Phys.}
{\bf 85}, 985 (1991); {\bf 86}, 599 (1991); {\bf 87}, 77 (1992);
N. Yamada, {\sl Sci. Rep. T\^ohoku Uni., Series 8}, {\bf 12}, 177
(1992); \pr {\bf A54}, 182 (1996); J.J.Halliwell and E.Zafiris,
{\sl Phys.Rev.} {\bf D57}, 3351-3364 (1998);
J.B.Hartle, {\sl Phys.Rev.} {\bf D44}, 3173 (1991);
%{\it Spacetime Coarse-Grainings in Non-Relativistic Quantum Mechanics.}
R.J.Micanek and J.B.Hartle, {\sl Phys.Rev.} {\bf A54},
3795 (1996).
% Nearly instantaneous alternatives in quantum mechanics.

\endreferences

\end